\documentclass[%
 aip,
 jcp,
 sd,%
 amsmath,amssymb,
preprint,%
]{revtex4-1}

\usepackage{color}
\usepackage{graphicx}
\usepackage{dcolumn}
\usepackage{bm}

\begin{document}


\title[]{Interaction and charge transfer between dielectric spheres: \\ exact and approximate analytical solutions}

\author{Fredrik Lind{\'e}n}
\author{Henrik Cederquist}%
\author{Henning Zettergren}%
 \email{henning@fysik.su.se.}
\affiliation{ Department of Physics, Stockholm University, SE-106 91 Stockholm, Sweden.
}%


\date{\today}

\begin{abstract}
We present exact analytical solutions for charge transfer reactions between two arbitrarily charged hard dielectric spheres. These solutions, and the corresponding exact ones for sphere-sphere interaction energies, include sums that describe polarization effects to infinite orders in the inverse of the distance between the sphere centers. In addition, we show that these exact solutions may be approximated by much simpler analytical expressions that are useful for many practical applications. This is exemplified through calculations of Langevin type cross sections for forming a compound system of two colliding spheres and through calculations of electron transfer cross sections. We find that it is important to account for dielectric properties and finite sphere sizes in such calculations, which for example may be useful for describing the evolution, growth, and dynamics of nanometer sized dielectric objects such as molecular clusters or dust grains in different environments including astrophysical ones.
%
\end{abstract}

\maketitle

\section{Introduction}


\footnotetext{\textit{$^{a}$~Address, Department of Physics, Stockholm University, SE-106 91 Stockholm, Sweden. Tel: +46 8 5537 8634; E-mail: henning@fysik.su.se}}

Accurate descriptions of the electrostatic interaction and charge transfer between particles of finite sizes is key to model a range of fundamental processes in science and engineering. A few examples here are cloud formation in the atmosphere \cite{Ochs:1987gsa}, like charge attraction in colloidal systems \cite{Bowen:1998ib},  stabilities of doubly charged fullerene clusters \cite{Huber:2016}, ion-mediated interactions and self-organization in nucleic acids and proteins \cite{Wong:2010kb}, dynamics of dusty plasmas\cite{Taccogna:2012fa}, collisional charging of interstellar dust grains \cite{Draine:1987va}, and toner particles in xerography photocopying techniques \cite{Feng:2000dc}.  

Various expressions for the interaction energy of a {\it point charge} and a polarizable neutral particle have been successfully used before to calculate Langevin type reaction rate constants for compound formation in gas-phase ion-neutral collisions in the meV to sub keV energy range (see e.g. Ref \cite{Eichelberger:2003hk} and references therein) and for electron transfer reactions involving fullerenes  \cite{Kasperovich200155} or nanoparticles \cite{Kasperovich:2000aa} . These studies \cite{Eichelberger:2003hk,Kasperovich200155,Kasperovich:2000aa} go beyond the original pioneering work of Langevin \cite{Langevin:1905wk}  who treated collisions between a point charge and a point-like neutral particle with finite polarizability $\alpha$. Further examples of extensions of the work of Langevin describe interactions between point charges and permanent electrical dipoles with random or fixed orientations \cite{Su:1992co,Hsieh:1981dn} or induced dipole - induced dipole interactions between a point charge with uniform \cite{Su:1978gg} or orientation-dependent\cite{Eichelberger:2003hk} polarizability  and a polarizable neutral. In the latter models \cite{Eichelberger:2003hk, Su:1978gg} the finite sizes of the collisions partners were indirectly taken into account by including polarizabilities and the number of outer shell electrons (HOMO electrons) as model parameters. This approach works well for small systems but for larger ones the actual geometrical sizes of both collision partners  are expected to become important \cite{Eichelberger:2003hk}. Draine and Sutin  \cite{Draine:1987va} considered the sizes of nanometer dust grains by treating them as metal spheres in interactions with point-charge ions.  In the present work, we calculate Langevin type reaction cross sections  (i.e. the cross sections for compound formation) for the interaction between two (charged) {\it dielectric} spheres and we present analytical expressions for electron (or charge) transfer between two dielectric spheres of finite sizes.  

For high energies ($\gtrsim$ 100 eV) the classical Langevin type reaction cross section is equal to the geometrical cross section $\pi$($a_A$+$a_B$)$^2$ of the two collision partners (hard sphere collisions) with radii $a_A$ and $a_B$. In this high energy regime, {\it charge transfer} cross sections are most often significantly larger than the geometrical cross section. The classical over-the-barrier model has been shown to accurately describe such processes in keV ion-atom \cite{Barany:1985cq,Niehaus:1986kg}, ion-fullerene\cite{Barany:1995hr,Cederquist:2000hv}, ion- PAH \cite{Lawicki:2011fc,Forsberg:2013cs}, fullerene-fullerene \cite{Zettergren:2002ja}, ion-surface \cite{Burgdorfer:1991gc} and fullerene-surface collisions\cite{Wethekam:2007be}. In these models \cite{Barany:1985cq,Niehaus:1986kg,Barany:1995hr,Cederquist:2000hv,Lawicki:2011fc,Forsberg:2013cs,Zettergren:2002ja,Burgdorfer:1991gc,Wethekam:2007be} , charge transfer takes place when the maximum of the potential energy barrier for the active electron equals the Stark shifted ionization energy of the target. For fullerene and PAH targets, analytical expressions for the potential energy barriers were derived assuming that these objects behave like metal spheres \cite{Barany:1995hr,Cederquist:2000hv} and infinitely thin metal discs \cite{Forsberg:2013cs}, respectively. These are good assumptions for isolated fullerenes and PAHs as they have been shown to display metallic behaviour when perturbed by external charges \cite{Zettergren:2012ft,Forsberg:2013cs}. However, weakly bound clusters of these species are expected to respond more like bulk material with dielectric constants in the $\epsilon$=5-7 range and it may thus be important to take finite $\epsilon$-values into account for more accurate descriptions of charge transfer processes in the general case.  

In this work we present exact analytical expressions for the potential energy for an electron in the presence of two spheres with arbitrary charges, radii, and dielectric constants. These expressions may then be used to calculate absolute charge-exchange cross sections.  It has previously been shown that the mutual polarization of two charged dielectric spheres in vacuum can be calculated using the image charge technique \cite{Lindell:1993jy}, from the surface charge densities \cite{Bichoutskaia:2010bf}, or with the aid of the so-called re-expansion method \cite{Nakajima:1999cr}.  In Sec. \ref{sec:model}, we use the latter method  \cite{Nakajima:1999cr} and derive the exact analytical expression also for the potential energy of a point charge outside two dielectric spheres. Further, we present approximate analytical expressions for the interaction energy of the two spheres and for the potential energy barrier for a point charge moving between the spheres, and demonstrate that these agree well with the corresponding exact analytical solutions. In Sec. \ref{sec:applications} we use these approximate expressions to calculate Langevin type compound formation (growth) cross sections and electron transfer cross sections between nanometer-sized spherical objects.

\section{Models for interaction and charge transfer between two dielectric spheres} \label{sec:model}

Here, we describe the electrostatic interaction between two charged dielectric spheres \cite{Nakajima:1999cr} and present the exact analytical solution for the interaction energy of a point charge in the presence of two dielectric spheres. We compare these solutions with much simpler approximate analytical expressions, which relate to an approximate expression for the electrostatic interaction between a point charge and a dielectric sphere (as shown in appendix A). 

\subsection{Two charged dielectric spheres}
\begin{figure}
\centering
\includegraphics[width=8cm,clip=, bb=0 0 350 120]{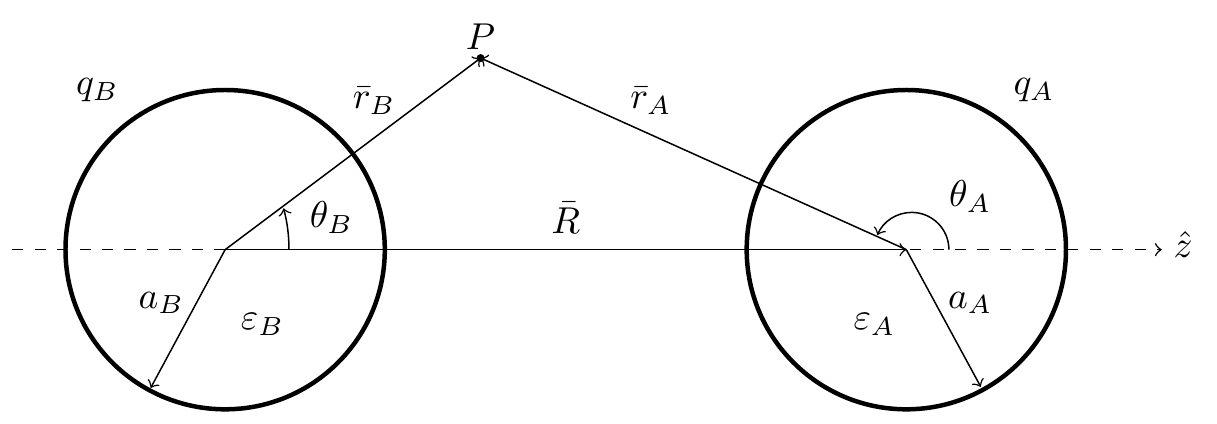}
\caption{Two coordinate systems (A and B) are used in the re-expansion method to describe the mutual electrostatic interaction between two charged dielectric spheres.}
\label{fig:dielectric_spheres}        
\end{figure}

We consider two dielectric spheres which are separated by the center-center distance $R$, and have charges $q_B$ and $q_A$, relative dielectric constants $\epsilon_B$ and $\epsilon_A$, and radii $a_B$ and $a_A$, respectively (see Fig. \ref{fig:dielectric_spheres}).  In the following, we use frequency independent $\epsilon_B$- and $\epsilon_A$-values to emphasize differences in reactivities between objects with different dielectric properties in the static limit. However, it is straightforward to introduce frequency dependencies by replacing $\epsilon_B$ and $\epsilon_A$ by $\epsilon_B (\omega)$ and $\epsilon_A(\omega)$  in the final expressions for the interaction energies and for charge transfer. Dynamic effects are indeed important for e.g. collisions between keV highly charged ions and insulator or semiconductor surfaces\cite{burgdorfer96,Hagg:1997aa}, where the typical interaction time scales are on the order of 10$^{-14}$ s. This is comparable to time scales in some of the problems that could be treated using the present formulas.

\subsubsection{The electrostatic potential outside two dielectric spheres. ~~}

The electrostatic potential at an arbitrary position P (see Fig. \ref{fig:dielectric_spheres}) outside the two dielectric spheres is \cite{Nakajima:1999cr}
\begin{eqnarray}\label{eq:Phi0_1}
\Phi_{out}(r_A,r_B,\theta _A,\theta _B)&=&\sum_{l=0}^{\infty}c_l^B\bigg{(}\frac{a_B}{r_B}\bigg{)}^{l+1}P_l(cos\theta _B) \nonumber \\
&+&\sum_{l=0}^{\infty}c_l^A\bigg{(}\frac{a_A}{r_A}\bigg{)}^{l+1}P_l(cos\theta _A),
\end{eqnarray}
where $r_B$ and $r_A$ are the distances from the centers of sphere $B$ and sphere $A$ to the point P, $\theta_B$ and $\theta_A$ are the angles from the axis connecting the sphere centers (z-axis), and $P_l$ is the $l^{th}$ order Legendre polynomial. Here we follow Nakajima and Sato \cite{Nakajima:1999cr} and use the re-expansion method to determine the Legendre coefficients $c_l^A$ and $c_l^B$ (for the explicit expressions of these coefficients see Appendix B).

\subsubsection{The interaction energy and force between two dielectric spheres. ~~}

The total electrostatic energy of the system consisting of two dielectric spheres is \cite{jackson75}
\begin{eqnarray}\label{eq:Uint-general}
U_{tot}(R)&=&\frac{1}{2}\int_0^{\pi}\sigma_B\Phi^B_{out}(a_B,\theta)2\pi a_B^2 sin\theta d\theta \nonumber\\
&+&\frac{1}{2}\int_0^{\pi}\sigma_A\Phi^A_{out}(a_A,\theta)2\pi a_A^2 sin\theta d\theta,  
\end{eqnarray}
where $\Phi^B_{out}(a_B,\theta)$ and $\Phi^A_{out}(a_A,\theta)$ are the surface potentials and  $\sigma_B$ and $\sigma_A$ are the surface charge densities on sphere B and sphere A, respectively, when they are at infinite separation, $R$, from each other. The latter, so-called free surface charge densities \cite{jackson75}, are given by
\begin{eqnarray}
\sigma_B&=&\frac{q_B}{4\pi a_B^2} \nonumber \\
\sigma_A&=&\frac{q_A}{4\pi a_A^2} \nonumber.
\end{eqnarray}
Using Eq. (\ref{eq:Phi0_1}) in Eq. (\ref{eq:Uint-general}) gives
\begin{eqnarray} 
U_{tot}(R)&=&\frac{q_A^2}{2a_A}+\frac{q_B^2}{2a_B}+\frac{q_Aq_B}{R} \nonumber \\
&+&\frac{1}{2}\sum_{l=1}^{\infty}\bigg{[}q_A\bigg{(}\frac{a_B}{R}\bigg{)}^{l+1}c_l^B-q_B\bigg{(}\frac{-a_A}{R}\bigg{)}^{l+1}c_l^A \bigg{]} \nonumber \\
\end{eqnarray}
where the sum of the first two terms corresponds to the self energies of the two spheres, i.e. the total energy of the system at infinite separation. The third term is the pure Coulomb energy and the infinite sum gives the shift in energy due to the mutual polarization which we denote $U_{pol}^B(R)$+$U_{pol}^A(R)$. The interaction energy for the two spheres is then
\begin{eqnarray}\label{eq:Uint-dielectric}
U_{int}(R)&=&U_{tot}(R)-\frac{q_A^2}{2a_A}-\frac{q_B^2}{2a_B}\nonumber \\
&=&\frac{q_Aq_B}{R}+ \nonumber \\
&&\frac{1}{2}\sum_{l=1}^{\infty}\bigg{[}q_A\bigg{(}\frac{a_B}{R}\bigg{)}^{l+1}c_l^B-q_B\bigg{(}\frac{-a_A}{R}\bigg{)}^{l+1}c_l^A \bigg{]} \nonumber \\
&=&\frac{q_Aq_B}{R}+U_{pol}^B(R)+U_{pol}^A(R).
\end{eqnarray}
Using the approximate expression for a point charge and a dielectric sphere  (Eq. \ref{eq:Uint_point_approx}) we arrive at an approximate expression for the interaction energy (Eq. (\ref{eq:Uint-dielectric})) between the two spheres,
\begin{eqnarray}\label{eq:Uint-dielectric_approx}
U_{int}(R)&\approx&\frac{q_Aq_B}{R} \nonumber \\
&-& \frac{\epsilon_{B}-1}{\epsilon_{B}+2}\frac{q_{A}^2a_{B}^3}{2R^2(R^2-a_{B}^2)}- \frac{\epsilon_{A}-1}{\epsilon_{A}+2}\frac{q_{B}^2a_{A}^3}{2R^2(R^2-a_{A}^2)}.\nonumber \\
\end{eqnarray}
This expression immediately indicates that the interaction energy is more sensitive to the charges and sizes of the collision partners than to the dielectric constants. However, for a given collision system (fixed charges and sizes) the dielectric constants may have large influences on the interaction energy, especially for $\epsilon$-values smaller than  $\sim$10.


\begin{figure}
\centering
\includegraphics[width=8cm,clip=, bb=0 0 450 350]{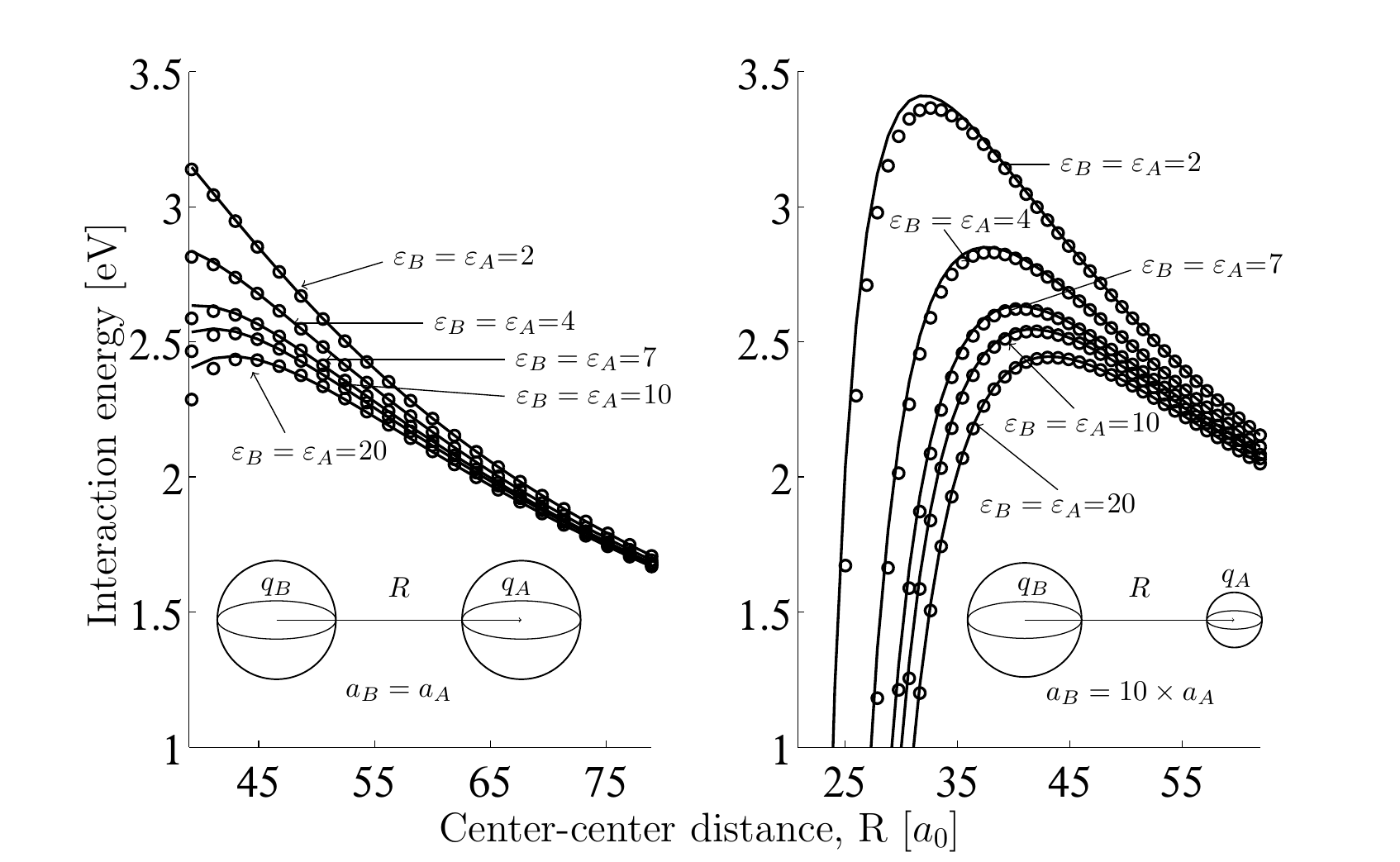}
\caption{The interaction energies as functions of the center-center distance $R$ between two charged dielectric spheres ($q_A$ = 5
and $q_B$= 1) for different values of the dielectric constant ($\epsilon_A$=$\epsilon_B$). The circles are the results from the exact analytical solution (Eq. \ref{eq:Uint-dielectric} ) and the solid lines are from the approximate expression (Eq. \ref{eq:Uint-dielectric_approx}). The sphere radii are set to 
$a_A$ = $a_B$ = 18.9 a$_0$ (1 nm) and to $a_A$ = 1.89 a$_0$ (0.1 nm) and $a_B$ = 18.9 a$_0$ (1 nm ) for the examples in the left and right panels, respectively.}
\label{fig:intE_spheres}        
\end{figure}

This is illustrated in Fig. \ref{fig:intE_spheres} where  we show comparisons between the approximate expression (Eq. \ref{eq:Uint-dielectric_approx}) and the exact analytical solution  (Eq. \ref{eq:Uint-dielectric} ) for the interaction energy $U_{int}(R)$. The left and right panels show results for two charged dielectric spheres ($q_A$ = 5 and $q_B$= 1) with equal radii of 1 nm ($a_A$ = $a_B$ = 18.9  a$_0$) 
and when one sphere radius is a factor of ten smaller than the other one ($a_A$ =1.89 a$_0$ and $a_B$ = 18.9  a$_0$), respectively.  The approximate expression yields results in close agreement with the exact analytical solution and may thus be used for practical applications with good performance at low computational costs. This has recently been demonstrated in studies of the stabilities of doubly charged fullerene clusters \cite{Huber:2016}, where Eq. \ref{eq:Uint-dielectric_approx} was successfully used to model the pair-wise interactions between neighboring fullerenes in the clusters. The approximate expression (Eq. \ref{eq:Uint-dielectric_approx})  may also be used to calculate e.g. kinetic energy releases of Coulomb exploding clusters \cite{Zettergren:2007ig}, the force between two charged spherical objects such as e.g. poly-methyl methacrylate (PMMA) spheres \cite{SunilKSainis:2007cp,Sainis:2008gj}, and Langevin reaction rates.

\subsection{Two charged dielectric spheres and a point charge}

\subsubsection{The potential energy barrier for charge transfer. ~~}

When a point charge $q_p$ is located at the position P in Fig. \ref{fig:dielectric_spheres}, it polarizes both (dielectric) spheres. The exact expression for the electrostatic potential at the position of the point charge is then given by   

\begin{eqnarray}\label{eq:Phi0_2}
\Phi_{out}(r_A, r_B, \theta _A,\theta _B)&=&\sum_{l=0}^{\infty}\bigg{(}c_l^B+\frac{\delta c_l^B}{2}\bigg{)}\bigg{(}\frac{a_B}{r_B}\bigg{)}^{l+1}P_l(cos\theta _B) \nonumber \\
&+&\sum_{l=0}^{\infty}\bigg{(}c_l^A+\frac{\delta c_l^A}{2}\bigg{)}\bigg{(}\frac{a_A}{r_A}\bigg{)}^{l+1}P_l(cos\theta _A)\nonumber \\
&&
\end{eqnarray}

where the coefficients $c_l^B$ and $c_l^A$  describe the contribution to the potential due to the mutual polarization of the two spheres (cf. Eq. \ref{eq:Phi0_1}). The $\delta c_l^B$ and $\delta c_l^A$ coefficients account for the polarization of the spheres by the point charge $q_P$, where the factor 1/2 origins from the self image force on the point charge. The latter contribution to the potential may be determined by applying the new boundary conditions conditions (with $q_P$ present in point P) and use the same procedure as for the pure sphere-sphere interaction (see Appendix C for details and explicit expressions for $\delta c_l^B$ and $\delta c_l^A$). 


The lowest potential energy barrier for a point charge $q_P$  in the presence of two dielectric spheres is always a saddle point located on the the axis connecting the sphere centers (the z-axis in Fig. 1). This is due to the azimuthal symmetry of the problem. The electrostatic potential for $q_P$ along the z-axis can be approximated by 
\begin{eqnarray}\label{eq:Barrier-dielectric_approx}
\Phi_{out}(z)&\approx&\frac{q_A}{R-z}+\frac{q_B}{z} \nonumber \\
&+& \frac{\epsilon_{B}-1}{\epsilon_{B}+2}\bigg{(}\frac{-q_{A}a_{B}^3}{Rz(Rz-a_{B}^2)}\bigg{)}\nonumber \\
&+& \frac{\epsilon_{A}-1}{\epsilon_{A}+2}\bigg{(}\frac{-q_{B}a_{A}^3}{R^2(R-z)^2-R(R-z)a_{A}^2}\bigg{)}\nonumber \\
&+& \frac{\epsilon_{B}-1}{\epsilon_{B}+2}\bigg{(}\frac{-q_{P}a_{B}^3}{2z^2(z^2-a_{B}^2)}\bigg{)}\nonumber \\
&+&\frac{\epsilon_{A}-1}{\epsilon_{A}+2}\bigg{(}\frac{-q_{P}a_{A}^3}{2((R-z)^4-(R-z)^2a_{A}^2)}\bigg{)}.
\end{eqnarray}
Eq. (\ref{eq:Barrier-dielectric_approx}) may thus be used for calculating the potential energy barrier for an electron moving between the two spheres.

\begin{figure}
\centering
\includegraphics[width=8cm,clip=, bb=0 0 320 200]{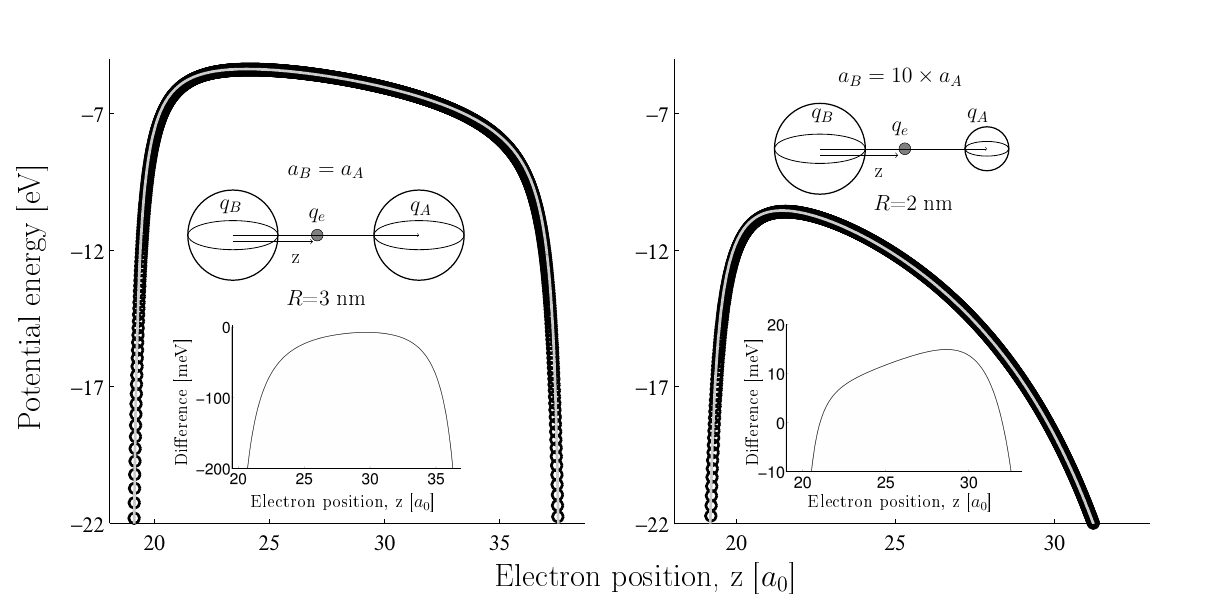}
\caption{The potential energy for an electron located between two charged dielectric spheres ($q_A$ = 5 and $q_B$ = 1) of equal dielectric constants ($\epsilon_A$=$\epsilon_B$=5). The circles are from the exact analytical solution (Eq. \ref{eq:Phi0_2} ) and the lines are from the approximate expression (Eq. \ref{eq:Barrier-dielectric_approx}). Left panel: Sphere radii 
$a_A$ = $a_B$ = 18.9  a$_0$ (1 nm), center-center distance $R$=56.7 a$_0$ (3 nm). Right panel: Sphere radii $a_A$ = 1.89  a$_0$  (0.1 nm) and $a_A$ = 18.9  a$_0$ (1 nm), center-center distance $R$=37.8 a$_0$ (2 nm). The insets shows the energy differences between the exact and the approximate solution. Note the meV scales in the insets.}
\label{fig:PE_spheres}        
\end{figure}

In Fig \ref{fig:PE_spheres} we show comparisons between the approximate (Eq. (\ref{eq:Barrier-dielectric_approx})) and exact analytical expressions (Eq. (\ref{eq:Phi0_2} )) for an electron ($q_p$=-1) located between two charged dielectric spheres ($q_A$ = 5, $q_B$ = 1, and $\epsilon_A$=$\epsilon_B$=5). The sphere radii are set to $a_A$ = $a_B$ = 1 nm = 18.9  a$_0$ (left panel) and to $a_A$ = 1.89  a$_0$  and $a_A$ = 18.9  a$_0$ (right panel). The approximate analytical expression gives results close to the exact analytical ones, especially in the region close to the maximum of the potential energy barrier (see the insets in Fig. \ref{fig:PE_spheres} where deviations between the exact and approximate solutions are shown in meV).
\begin{figure}
\centering
\includegraphics[width=8cm,clip=, bb=0 0 420 300]{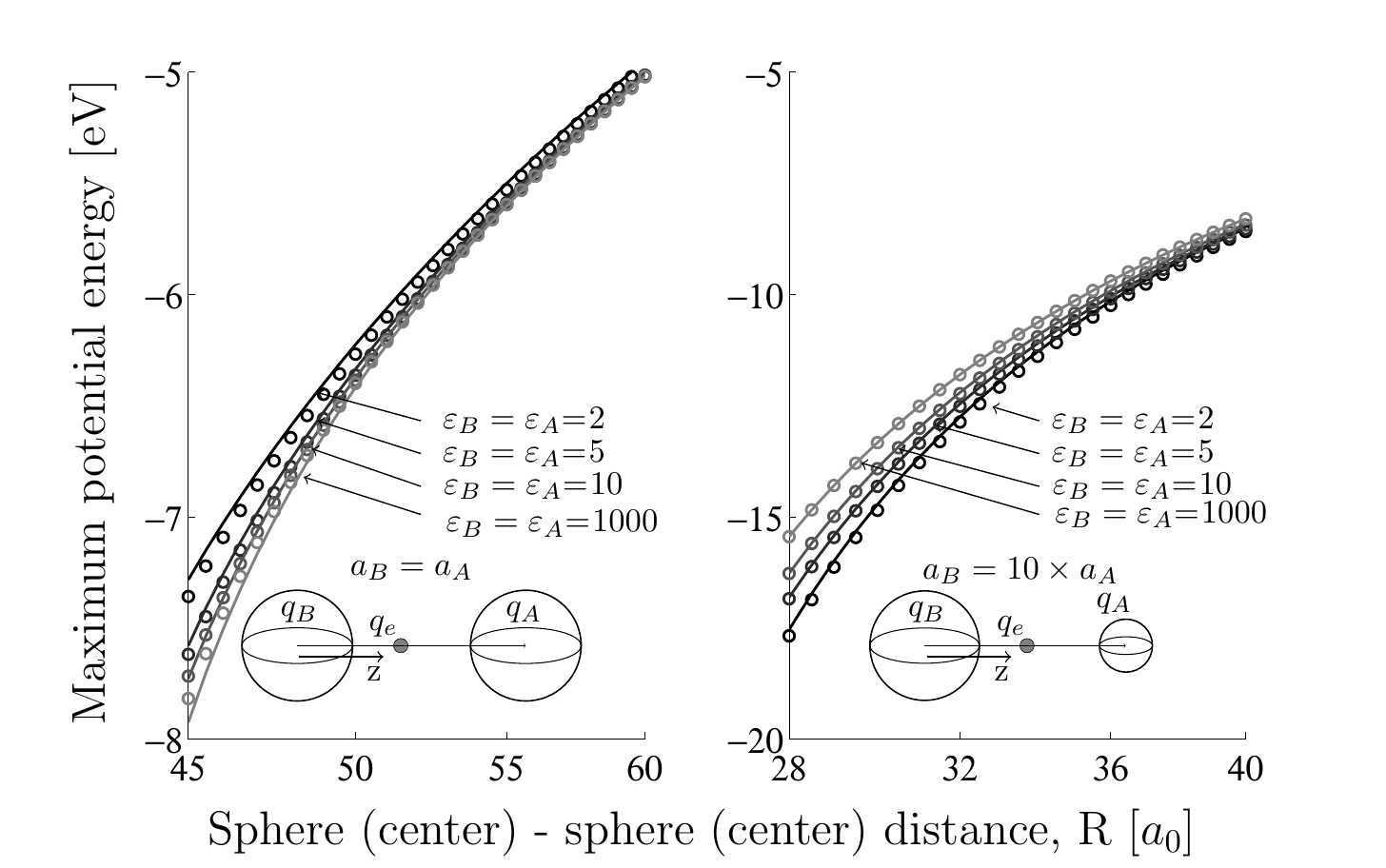}
\caption{Maxima of the potential energy barriers for an electron located on the axis connecting two charged dielectric spheres ($q_A$ = 5 and $q_B$ = 1, $\epsilon_A$=$\epsilon_B$=5) as functions of the center-center distance $R$. The circles are the results from the exact analytical solution (Eq. \ref{eq:Phi0_2} ) and the solid lines are those from the approximate expression (Eq. \ref{eq:Barrier-dielectric_approx}). Left panel: Sphere radii  $a_A$ = $a_B$ = 18.9  a$_0$ (1 nm) . Right panel: Sphere radii $a_A$ = 1.89  a$_0$  (0.1 nm) and $a_A$ =  18.9  a$_0$ (1 nm).}
\label{fig:PEmax_spheres}        
\end{figure}
This is further illustrated in Fig \ref{fig:PEmax_spheres}, where we show the maximum of the potential energy barriers as functions of the center-center distance for a range of dielectric constants (the same for both spheres). 

\subsubsection{Stark-shifted electron binding energies~}

Adopting the classical over-the-barrier concept, an electron is classically allowed to transfer from the target sphere to the projectile sphere at distances $R$ (see Fig. \ref{fig:dielectric_spheres} ) where the maximum of the potential energy barrier ($V$=$q_p$$\Phi_{out}$=$-\Phi_{out}$) for the active electron is lower than its Stark shifted binding energy
\begin{eqnarray}\label{eq:Starkshift1}
I_1^*&=&I_1+\Phi_{Stark}(R=R_{1})
\end{eqnarray}
to the target. Here $\Phi_{Stark}$ refers to the Stark shift at the critical distance, $R$=$R_1$, where electron transfer first becomes classically allowed. $I_1$ is the binding energy of the electron to the sphere with charge $q_B$=1 in the absence of the sphere charge $q_A$ (i.e. with $q_A$ at $R$=$\infty$). The Stark-shifted binding energy of the $n$th electron to the sphere of charge $q_B$=$n$+1 at the critical distance $R$=$R_{n+1}$ for the transfer of the ($n$+1)st electron from the sphere is 
\begin{equation}\label{eq:Starkshift}
I_{n+1}^*=I_{n+1}+\Phi_{Stark}(R=R_{n+1})
\end{equation}
where the point charge now is reduced to $q_A$-$n$. That is we follow B\'ar\'any {\it et al} \cite{Barany:1985cq} and Hvelplund {\it et al} \cite{hvelplund1985energy}  and assume sequential electron transfer and that previously transferred electrons fully screen the charge $q_A$. 

The Stark shift term  $\Phi_{Stark}$  for removing an electron from a conducting sphere is simply $q_A$/$R$ (the surface potential).  For charge transfer between two dielectric spheres, Eq. (\ref{eq:Phi0_1}) gives the exact surface potential $q_P$$\Phi_{out}$($r_B$=$a_B$, $\theta_B$)=-$\Phi_{out}$($r_B$=$a_B$, $\theta_B$) for an infinite number of terms in the Legendre expansion. Alternatively, one can use the approximate expression

\begin{eqnarray}\label{eq:Surf_approx}
\Phi_{out}^{B}(r_B=a_B, \theta_B) \approx \frac{q_{A}}{\sqrt{R^2+a_B^2-2a_BRcos\theta_B}} + &\nonumber \\
\frac{\epsilon_{B}-1}{\epsilon_{B}+2}\bigg{(}\frac{q_{A}}{R}-\frac{q_{A}}{\sqrt{R^2+a_B^2-2a_BRcos\theta_B}}\bigg{)}&,\nonumber \\
\end{eqnarray}

which gives the correct asymptotic limit when $\epsilon_B \rightarrow \infty$, $\Phi_{out}^{B}$($r_B$=$a_B$, $\theta_B$) = $q_A$/$R$.

We find that the approximate expression for the surface potential (Eq. \ref{eq:Surf_approx}) does not deviate significantly from the exact analytical expression for finite values of the dielectric constants and the radius of the sphere ($a_B$).   The surface potential typically displays a rather weak dependence of $\theta_B$ close to the point where the z-axis (connecting the sphere centers) crosses the sphere surface. Here, we thus assume that the Stark shift is given by the potential experienced by an active electron at the position on the surface closest to the projectile ($r_B$=$a_B$, $\theta_B$=0)
\begin{eqnarray}\label{eq:Starkshift_approx}
\Phi_{Stark}&=&\Phi_{out}^{B}(r_B=a_B, \theta_B=0) \nonumber \\
&\approx& \frac{q_{A}}{R-a_B} + \frac{\epsilon_{B}-1}{\epsilon_{B}+2}\bigg{(}\frac{q_{A}}{R}-\frac{q_{A}}{R-a_B}\bigg{)}.\nonumber \\
\end{eqnarray}
We thus follow a similar approach as has been used for ion-surface collisions\cite{burgdorfer96} where the shifted target levels are $\Phi_{Stark}(a_B \rightarrow \infty)$=2$q_A$/(D($\epsilon_{B}$+1)) for an ion-surface distance $D$.

\section{Model results} \label{sec:applications}

\subsection{Reaction cross sections in collisions between two charged spherical objects.}

In Fig. \ref{fig:Langevin} we show Langevin type reaction cross sections for compound formation (see Appendix D) as functions of the center of mass energy in collisions between two dielectric spheres with $q_A$=1 and $q_B$=-1 (triangles), $q_A$=1 and $q_B$=0 (squares), and $q_A$=1 and $q_B$=1 (circles). We compare two different values of the dielectric constants $\epsilon_A$=$\epsilon_B$=1000 (open symbols) and $\epsilon_A$=$\epsilon_B$=5 (full symbols) for three different sphere radii combinations. The sphere radii are, from top to bottom, $a_A$ = 0.1 nm and $a_B$= 1 nm,  $a_A$ = 1 nm and $a_B$= 1 nm, and $a_A$ = 1 nm and $a_B$= 10 nm. The overall picture is similar in all three panels. For high collision energies ($\gtrsim$ 10 eV) the reaction cross sections approach the geometrical cross sections $\pi$($a_A$+$a_B$)$^2$. For lower energies the attractive Coulomb and induced polarisation forces lead to a dramatic increase in the cross sections for oppositely charged spheres. The trends are similar  for interactions with neutral spheres but weaker since the attraction is due to induced polarisation effects only.  For like charges the Coulomb repulsion leads to decreasing cross sections for low energies. Note, however, that the cross sections for like charges (here $q_A$=1, $q_B$=1) have maxima when the sphere radii are different (lower and upper panels of Fig.  \ref{fig:Langevin}) and are larger than the geometrical cross sections for some ranges of the collision energies. The reason for this is the delicate balance between the Coulomb repulsion, the attractive mutual polarisation, and the centrifugal terms which governs the height of the reaction barrier (see Appendix D). 

The examples in Fig. \ref{fig:Langevin} show that the radii and charges have large influence on the reaction cross sections (see the different y-scales). The dielectric constant only plays a relatively small role in these sphere-size and charge-state regimes, but becomes important for higher charge states.

\begin{figure}
\centering
\includegraphics[width=7cm,clip=,, bb=0 0 400 700]{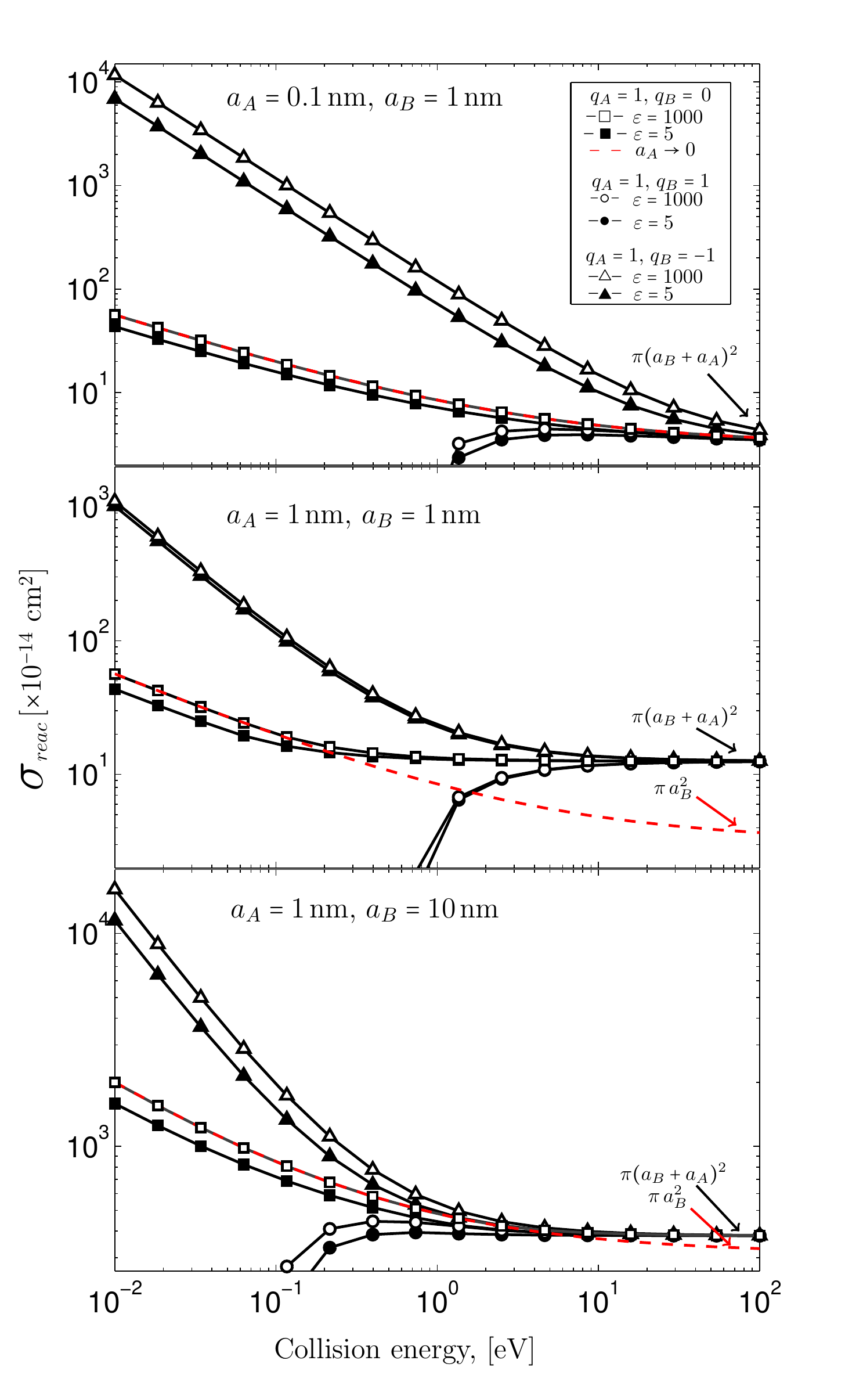}
\caption{Langevin type cross sections for compound formation in collisions between a positively charged and a negatively charged dielectric sphere (triangles), a positively charged  and a neutral dielectric sphere (squares), and two positively charged dielectric spheres (circles). The results were obtained using the approximate expression for the interaction energy (Eq. \ref{eq:Uint-dielectric_approx}). The dielectric constants are set to $\epsilon_A$=$\epsilon_B$=1000 (open symbols) and to $\epsilon_A$=$\epsilon_B$=5 (full symbols). The sphere radii are $a_A$ = 0.1 nm and $a_B$=1 nm (top panel), $a_A$ = 1 nm and $a_B$=1 nm (middle panel), and $a_A$ = 1 nm and $a_B$=10 nm (bottom panel). The dashed curves show the corresponding cross sections for a point charge interacting with a neutral metal sphere of radius $a_B$ (Eq. \ref{eq:langevin_cross_approx}).}
\label{fig:Langevin}        
\end{figure}

\subsection{Absolute electron transfer cross sections in collisions between two charged spherical objects.}
\begin{figure}
\centering
\includegraphics[width=8cm,clip=, bb=0 0 440 550]{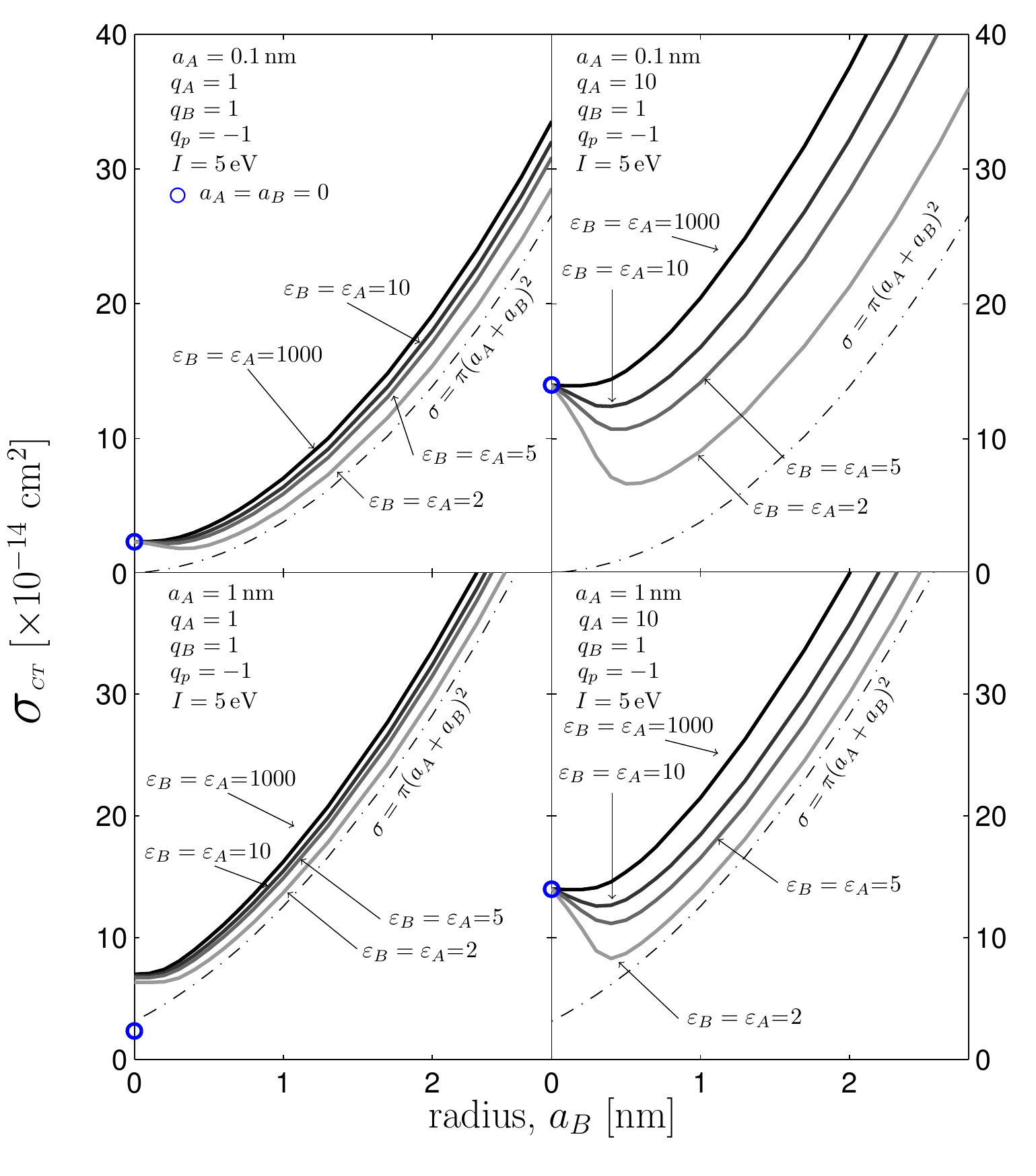}
\caption{Absolute electron transfer cross sections for collisions between a charged dielectric sphere ($q_A$) and a neutral dielectric sphere ($q_B$ +$q_P$=1-1=0). These are shown as functions of the radius of the neutral sphere ($a_B$) and are calculated with the approximate expressions for the potential energy barrier (Eq. \ref{eq:Barrier-dielectric_approx}) and the Stark shifted ionization energy (Eq. \ref{eq:Starkshift_approx}) for different values of the dielectric constants  ($\epsilon_A$=$\epsilon_B$=2, 5, 10 and 1000).  The projectile charge and radius are $q_A$=1 and $a_A$=0.1 nm (upper left),  $q_A$=1 and $a_A$=1 nm (lower left),  $q_A$=10 and $a_A$=0.1 nm (upper right) and  $q_A$=10 and $a_A$=1 nm (lower right). The ionization energy of the neutral sphere is set to $I$=5 eV in all cases. The blue circles are the cross sections for collisions between two point charges ($a_A$=$a_B$=0).}
\label{fig:Cross_secs}        
\end{figure}

In Fig. \ref{fig:Cross_secs} we show the absolute electron transfer cross sections for collisions between a charged dielectric sphere ($q_A$) and a neutral dielectric sphere ($q_B$=1 and $q_P$=-1). The ionization energy is set to $I$= 5 eV in this example. The four panels show the cross sections for different values of the projectile charge ($q_A$) and radius ($a_A$)  as functions of the radius of the neutral sphere ($a_B$). Each panel include results for four different values of the dielectric constants ($\epsilon_A$=$\epsilon_B$=2, 5, 10 and 1000). In the upper left panel,  we show results for a singly charged small projectile ($q_A$=1, $a_A$=0.1 nm). Here the low charge state means that the projectile needs to come close to the target surface to capture an electron. The cross section is therefore close to the geometrical one $\pi$($a_A$+$a_B$)$^2$ and only weakly dependent on the dielectric constant. The same is true when the projectile radius is ten times larger (see the lower left panel of Fig.  \ref{fig:Cross_secs}). Note, however, that it is important to take the finite sizes of the collision partners into account as the present model values are significantly higher than those predicted by the classical over-the-barrier model for two point charges (indicated by $a_A$=$a_B$=0 in Fig. \ref{fig:Cross_secs}). 

In the upper right panel of Fig.  \ref{fig:Cross_secs}, we show results for a small ten times charged projectile ($q_A$=10, $a_A$=0.1 nm). Here, electron capture takes place far away from the sphere surface and thus the model cross sections are significantly larger than the geometrical cross section. Here, the polarization effects are much stronger than for singly charged projectiles (see the left panels of Fig.  \ref{fig:Cross_secs})) and the cross section is therefore more sensitive to the dielectric constant.  Note that the cross sections are significantly larger than the ones for two point charges (indicated by open blue circles) when the target sphere radius is large. Similar trends are seen when the projectile radius is increased by a factor of ten (see the lower right panel of Fig.  \ref{fig:Cross_secs}).  The results in Fig.  \ref{fig:Cross_secs} clearly illustrate the importance of considering the finite sizes and the dielectric constants in interactions between two (spherical) objects.

\section{Summary and conclusions}

In summary, we have derived an exact analytical solution for the potential energy experienced by a point charge moving between two charged dielectric spheres of arbitrary radii, charges, and dielectric constants. We have shown that this solution and the one describing the interaction energy for two charged dielectric spheres \cite{Nakajima:1999cr} may be well-approximated by simple analytical expressions which may be evaluated at very low computational costs. The latter expressions may thus be efficiently used for calculations of Langevin type rates for compound formation and electron transfer rates in collisions between two (charged) spherical objects, where the polarization (finite sizes) of  {\it both} collision partners are taken into account. Such rates have been reported for e.g. {\it point-charge} ions interacting with dust grains \cite{Draine:1987va, weingartner2001electron} and fullerenes \cite{oomont2016interstellar} . The expressions given in the present work may thus be used to gauge the significance of interactions between two finite size objects (e.g. molecules, clusters, and dust grains) in, for example, astrophysical environments.

\begin{acknowledgments} This work was supported by the Swedish Research Council (Contracts Nos. 621-2012-3662, 621-2012-3660, and 2015-04990). We acknowledge the COST action CM1204 XUV/X-ray Light and Fast Ions for Ultrafast Chemistry (XLIC).
\end{acknowledgments}


\appendix

\section{Interaction energy and force between one point charge and one dielectric sphere} \label{sec:model-dielectric_sphere}

\begin{figure}
\centering
\includegraphics[width=6cm,clip=,bb=0 0 350 250]{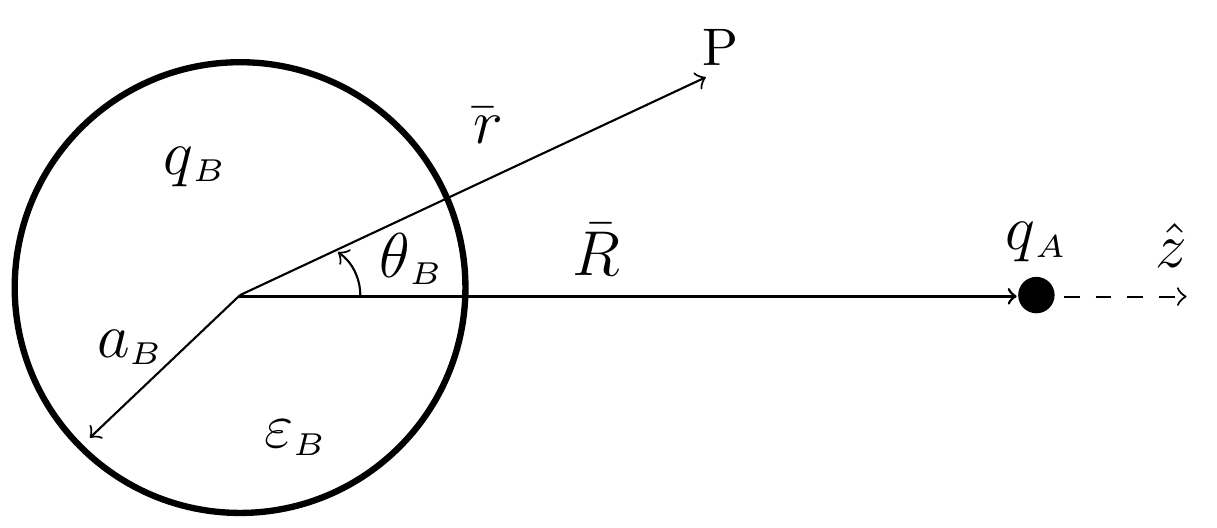}
\caption{Parameters for calculations of electrostatic potentials at points P in the presence of a point charge $q_A$ and a sphere with radius $a_B$, charge $q_B$, and relative dielectric constant $\epsilon_B$ . The positions of $q_A$ and P are defined by the vectors  ${\bar R}$ and  ${\bar r}$. The angle between  ${\bar R}$ and  ${\bar r}$ is $\theta_B$ .} 
\label{fig:dielectric_point_sphere}        
\end{figure}


Here, we consider a dielectric sphere with charge $q_B$ relative dielectric constant $\epsilon_B$ and radius $a_B$ in vacuum. A point charge $q_A$ is located at the position ${\bar R}$ (Fig. \ref{fig:dielectric_point_sphere}).
%
%
The force on this point charge is given by the text book formula \cite{franklin05}
\begin{eqnarray}\label{eq:Uint_point}
F(R)&=&-\frac{dU_{int}(R)}{dR} \nonumber \\
&=&\frac{q_Aq_B}{R^2}-\frac{q_A^2}{R^2}\sum_{l=1}^{\infty}\frac{l(l+1)(\epsilon_{B}-1)}{[l(\epsilon_{B}+1)+1]}\frac{a_B^{2l+1}}{R^{2l+1}},
\end{eqnarray}
where $U_{int}$ is the sphere point-charge interaction energy, 
\begin{equation}\label{eq:Uint_point}
U_{int}(R)=\frac{q_Aq_B}{R}-\frac{q_A^2}{2R}\sum_{l=1}^{\infty}\frac{l(\epsilon_{B}-1)}{[l(\epsilon_{B}+1)+1]}\frac{a_B^{2l+1}}{R^{2l+1}}.
\end{equation}
This energy is equal to the work required to move $q_A$ from the distance $R$=$|{\bar R}|$ to infinite separation \cite{Zettergren:2012ft}. Note that the induced charge distribution on the sphere changes with $R$ as the point charge is moved to infinity ($U_{int}$=$\int^{\infty}_R F(R) dr$), which gives the factor 1/2 in front of the infinite sum (see  e.g. Ref. \cite{Zettergren:2012ft} for details). 
When we let $\epsilon_{B} \rightarrow \infty$, Eq. (\ref{eq:Uint_point}) reduces to the interaction energy for a point charge and a conducting sphere \cite{Cederquist:2000hv, Zettergren:2012ft}
\begin{equation}\label{eq:Uint_point_met}
\lim\limits_{\epsilon_{B} \to \infty} U_{int}(R)=\frac{q_Aq_B}{R}-\frac{1}{2}\bigg{(}\frac{q_A^2a_B}{R^2-a_B^2}-\frac{q_A^2a_B}{R^2}\bigg{)}.
\end{equation}
Draine and Sutin \cite{Draine:1987va} showed that the exact expression for the interaction energy for a point charge and a dielectric sphere (Eq. (\ref{eq:Uint_point}) ) may be approximated by multiplying the second term in the Eq. (\ref{eq:Uint_point_met}) with the factor $(\epsilon_{B}-1)$/$(\epsilon_{B}+2)$, which gives 
\begin{eqnarray}\label{eq:Uint_point_approx}
U_{int}(R) &\approx& \frac{q_Aq_B}{R}-\frac{\epsilon_{B}-1}{2(\epsilon_{B}+2)}\bigg{(}\frac{q_A^2a_B}{R^2-a_B^2}-\frac{q_A^2a_B}{R^2}\bigg{)} \nonumber \\
&=&\frac{q_Aq_B}{R}- \frac{\epsilon_{B}-1}{\epsilon_{B}+2}\frac{q_{A}^2a_{B}^3}{2R^2(R^2-a_{B}^2)}.
\end{eqnarray}

\begin{figure}
\centering
\includegraphics[width=8cm,clip=, bb=0 0 370 300]{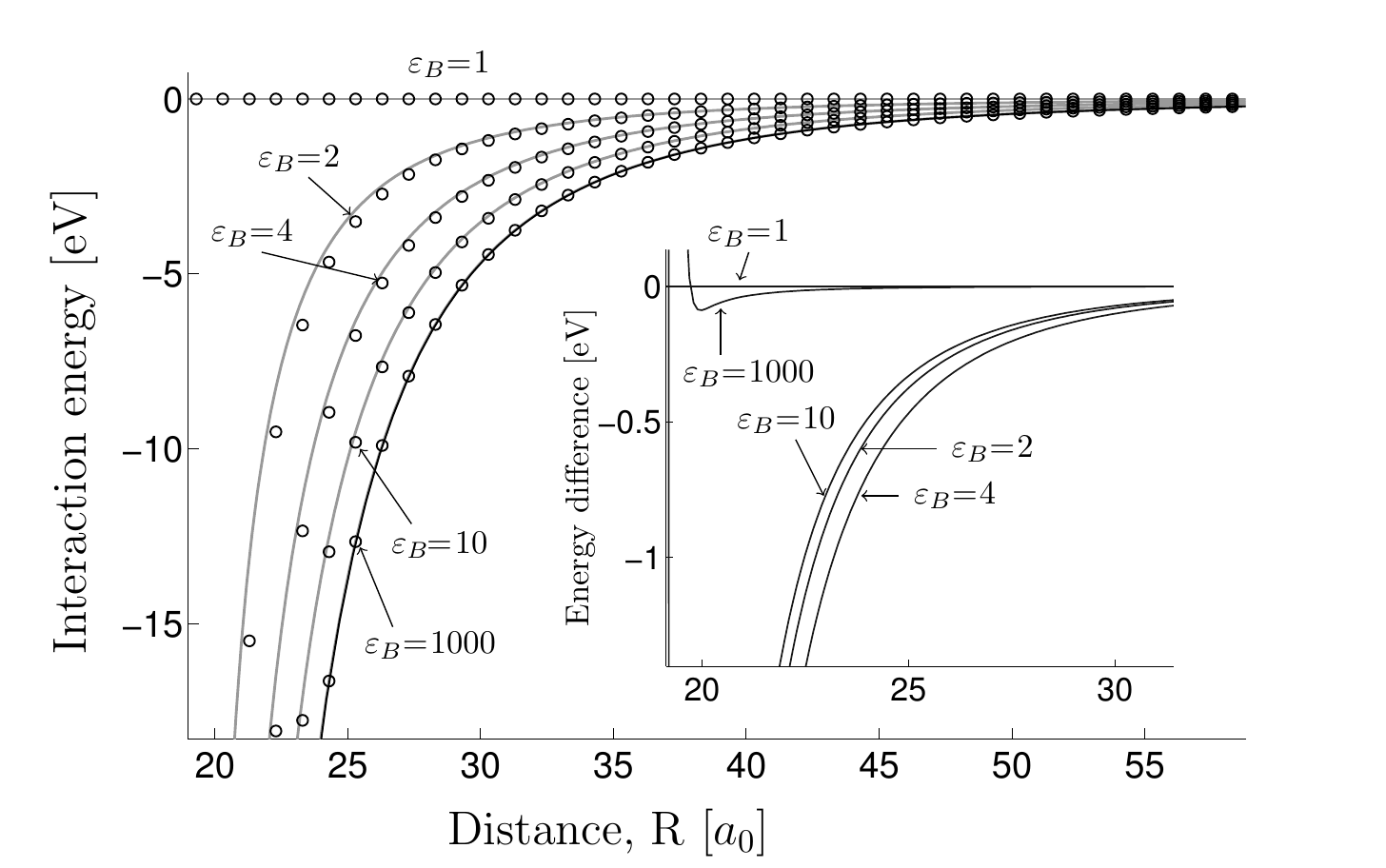}
\caption{ The electrostatic interaction energy $U_{int}(R)$ as a function of the distance, $R$, between a point charge ($q_A$=5) and a neutral dielectric sphere ($q_B$=0) with radius a$_B$= 18.9 a$_0$ (1 nm) and different values of the dielectric constant ($\epsilon_B$). The circles are the results using the exact analytical solution for $U_{int}(R)$ (Eq. \ref{eq:Uint_point}) and the solid lines are from the approximate expression (Eq. \ref{eq:Uint_point_approx}). The inset shows the energy difference between these expressions.} 
\label{fig:intE_point_sphere}        
\end{figure}

In Fig. \ref{fig:intE_point_sphere} we show comparisons between the exact (Eq. \ref{eq:Uint_point}) and the approximate expression (Eq. \ref{eq:Uint_point_approx}) for an example with a point charge $q_A$=5 interacting with a neutral dielectric sphere $q_B$=0 of radius a$_B$= 18.9 a$_0$ (1 nm). The exact and approximate results are in good agreement, especially at large separations $R$ which are important when calculating e.g. Langevin reaction rates. The approximate expression in Eq. \ref{eq:Uint_point_approx} may thus be efficiently used to calculate such rates in collisions between a point charge and a polarizable spherical object with dielectric constant $\epsilon_{B}$ \cite{Draine:1987va}.

\section{The Legendre polynomial coefficients for the interaction between two charged dielectric spheres}
According to the re-expansion method \cite{Nakajima:1999cr}, the electrostatic potential outside two charged dielectric spheres (Eq. (\ref{eq:Phi0_1})) may be expressed in coordinate systems B and A  (see Fig. 4 for the definitions of $r_A$, $r_B$, $a_A$, $a_B$, $\theta_A$, $\theta_B$, and $R$) as
\begin{equation}\label{eq:Phi0_B}
\Phi_{out}^B=\sum_{l=0}^{\infty}\bigg{[}c_l^B\bigg{(}\frac{a_B}{r_B}\bigg{)}^{l+1}+d_l^B\bigg{(}\frac{r_B}{a_B}\bigg{)}^{l}\bigg{]}P_l(cos\theta _B) 
\end{equation}
or as
\begin{equation}\label{eq:Phi0_A}
\Phi_{out}^A=\sum_{l=0}^{\infty}\bigg{[}c_l^A\bigg{(}\frac{a_A}{r_A}\bigg{)}^{l+1}+d_l^A\bigg{(}\frac{r_A}{a_A}\bigg{)}^{l}\bigg{]}P_l(cos\theta _A). 
\end{equation}
Here, the coefficients $d_l^B$ and $d_l^A$ are given by 
\begin{eqnarray} \label{eq:dlr}
d_l^B&=&-\bigg{(}\frac{a_B}{R}\bigg{)}^{l}\sum_{m=0}^{\infty}\frac{(m+l)!}{m!l!}\bigg{(}-\frac{a_A}{R}\bigg{)}^{m+1}c_m^A \\ \label{eq:dlq}
d_l^A&=&\bigg{(}-\frac{a_A}{R}\bigg{)}^{l}\sum_{m=0}^{\infty}\frac{(m+l)!}{m!l!}\bigg{(}\frac{a_B}{R}\bigg{)}^{m+1}c_m^B 
\end{eqnarray}
Note that these relations are independent of the boundary conditions. It is also important to remember that Eq. (\ref{eq:Phi0_1}) should be used to calculate the potential outside the spheres (not Eqs. (\ref{eq:Phi0_B}) and (\ref{eq:Phi0_A})) for convergence reasons \cite{Nakajima:1999cr}.

In order to determine the potential inside the spheres, the boundary conditions have to be taken into account. The surface charge densities ($\sigma_B$ and $\sigma_A$) are uniform when the separation between spheres A and B is infinite ($R=\infty$),
\begin{eqnarray}
\sigma_B&=&\frac{q_B}{4\pi a_B^2} \nonumber \\
\sigma_A&=&\frac{q_A}{4\pi a_A^2} \nonumber,
\end{eqnarray}
which thus correspond to the free surface charge densities \cite{jackson75}. Since there are no such free charges inside the spheres, the potential is given by 
\begin{equation}\label{eq:Phi_in_B}
\Phi_{in}^B=\sum_{l=0}^{\infty}e_l^B\bigg{(}\frac{r_B}{a_B}\bigg{)}^{l}P_l(cos\theta _B)  
\end{equation}
and
\begin{equation}
\Phi_{in}^A=\sum_{l=0}^{\infty}e_l^A\bigg{(}\frac{r_A}{a_A}\bigg{)}^{l}P_l(cos\theta _A).
\end{equation}
On the surface of the spheres ($r_B=a_B$ and $r_A=a_A$), the normal and tangential boundary conditions for the $\overline{D}$-   and the $\overline{E}$ -fields have to be fullfilled,
\begin{eqnarray}
\epsilon_B\frac{\partial \Phi_{in}^B}{\partial r_B}\bigg{|}_{r_B=a_B}-\frac{\partial \Phi_{out}^B}{\partial r_B}\bigg{|}_{r_B=a_B}&=&4\pi\sigma_BP_0(cos\theta_B)  \nonumber \\
\epsilon_A\frac{\partial \Phi_{in}^A}{\partial r_A}\bigg{|}_{r_A=a_A}-\frac{\partial \Phi_{out}^A}{\partial r_A}\bigg{|}_{r_A=a_A}&=&4\pi\sigma_AP_0(cos\theta_A) \nonumber
\end{eqnarray}
and
\begin{eqnarray}
-\frac{1}{a_B}\frac{\partial \Phi_{in}^B}{\partial \theta_B}\bigg{|}_{r_B=a_B}&=&-\frac{1}{a_B}\frac{\partial \Phi_{out}^B}{\partial \theta_B}\bigg{|}_{r_B=a_B} \nonumber\\
-\frac{1}{a_A}\frac{\partial \Phi_{in}^A}{\partial \theta_A}\bigg{|}_{r_A=a_A}&=&-\frac{1}{a_A}\frac{\partial \Phi_{out}^A}{\partial \theta_A}\bigg{|}_{r_A=a_A}. \nonumber
\end{eqnarray} 
This gives the following relations between the coefficients,
\begin{eqnarray}\label{eq:clr}
d_l^B=-\frac{l(\epsilon_B+1)+1}{l(\epsilon_B-1)}c_l^B &, & e_l^B=-\frac{2l+1}{l(\epsilon_B-1)}c_l^B \nonumber \\
c_0^B=\frac{q_B}{a_B}&, & e_0^B=c_0^B+d_0^B 
\end{eqnarray}
and
\begin{eqnarray}\label{eq:clq}
d_l^A=-\frac{l(\epsilon_A+1)+1}{l(\epsilon_A-1)}c_l^A &, & e_l^A=-\frac{2l+1}{l(\epsilon_A-1)}c_l^A	\nonumber \\
c_0^A=\frac{q_A}{a_A}&, & e_0^A=c_0^A+d_0^A 
\end{eqnarray}
in sphere B and A, respectively. If we now combine the results in Eqs. (\ref{eq:dlr}), (\ref{eq:dlq}), (\ref{eq:clr}), and (\ref{eq:clq}), we arrive at 
\begin{eqnarray} \label{eq:clB}
c_l^B&=&\frac{l(\epsilon_B-1)}{l(\epsilon_B+1)+1}\bigg{(}\frac{a_B}{R}\bigg{)}^l \nonumber \\
&&\times\bigg{[}\frac{-q_A}{ R} +\sum_{m=1}^{\infty}\bigg{(}\frac{-a_A}{R}\bigg{)}^{m+1}\frac{(m+l)!}{m!l!}c_m^A\bigg{]} \\
\label{eq:clA}
c_l^A&=&-\frac{l(\epsilon_A-1)}{l(\epsilon_A+1)+1}\bigg{(}\frac{-a_A}{R}\bigg{)}^l \nonumber \\
&&\times\bigg{[}\frac{q_B}{R} +\sum_{m=1}^{\infty}\bigg{(}\frac{a_B}{R}\bigg{)}^{m+1}\frac{(m+l)!}{m!l!}c_m^B\bigg{]} 
\end{eqnarray}
for $l>0$. After some algebraic exercise, these equations can be written in a more compact form 
 \begin{eqnarray}\label{eq:h_B}
\bf (h^B-I)c^B&=&\bf w^B \\ \label{eq:h_A}
\bf (h^A-I)c^A&=&\bf w^A,
\end{eqnarray}
where $\bf I$ is the unit matrix and $\bf c^B$ and $\bf c^A$ are two column vectors with the elements $c_l^B$ and
 $c_l^A$, respectively. The matrix and vector elements are given by
 
\begin{eqnarray} h^B_{l,m}&=&f(\epsilon_B,l)\bigg{(}\frac{a_B}{R}\bigg{)}^{l+m+1}\bigg{(}\frac{a_A}{R}\bigg{)} \nonumber \\
&\times& \sum_{j=1}^{\infty}f(\epsilon_A,j)\frac{(j+l)!}{j!l!}\frac{(j+m)!}{j!m!}\bigg{(}\frac{a_A}{R}\bigg{)}^{2j} 
\label{eq:H_B} \\
h^A_{l,m}&=&-f(\epsilon_A,l)\bigg{(}\frac{-a_A}{R}\bigg{)}^{l+m+1}\bigg{(}\frac{a_B}{R}\bigg{)} \nonumber\\
&\times& \sum_{j=1}^{\infty}f(\epsilon_B,j)\frac{(j+l)!}{j!l!}\frac{(j+m)!}{j!m!}\bigg{(}\frac{a_B}{R}\bigg{)}^{2j} 
\label{eq:H_A} \\
w^B_{l}&=&f(\epsilon_B,l)\frac{1}{R}\bigg{(}\frac{a_B}{R}\bigg{)}^{l} \nonumber\\
&\times& \bigg{[}q_A-q_B\bigg{(}\frac{a_A}{R}\bigg{)}\sum_{j=1}^{\infty}f(\epsilon_A,j)\frac{(j+l)!}{j!l!}\bigg{(}\frac{a_A}{R}\bigg{)}^{2j}\bigg{]} 
\nonumber\\
\\
w^A_{l}&=&f(\epsilon_A,l)\frac{1}{R}\bigg{(}\frac{-a_A}{R}\bigg{)}^{l} \nonumber\\
&\times& \bigg{[}q_B-q_A\bigg{(}\frac{a_B}{R}\bigg{)}\sum_{j=1}^{\infty}f(\epsilon_B,j)\frac{(j+l)!}{j!l!}\bigg{(}\frac{a_B}{R}\bigg{)}^{2j}\bigg{]} 
\nonumber\\
\end{eqnarray}
where 
\begin{equation}\label{eq:f}
f(\epsilon,j)=\frac{j(\epsilon-1)}{j(\epsilon+1)+1}. 
\end{equation}

\section{The Legendre polynomial coefficients for a point charge between the two charged dielectric spheres}

The additional Legendre polynomial coefficients due to the presence of a point charge are 
\begin{eqnarray}
\delta c_l^B&=&\frac{l(\epsilon_B-1)}{l(\epsilon_B+1)+1}\bigg{[}\frac{-q_p}{ r}\bigg{(}\frac{a_B}{r}\bigg{)}^l 
\nonumber \\
&&+\bigg{(}\frac{a_B}{R}\bigg{)}^l\sum_{m=1}^{\infty}\bigg{(}\frac{-a_A}{R}\bigg{)}^{m+1}\frac{(m+l)!}{m!l!}\delta c_m^A\bigg{]} \\
\delta c_l^A&=&-\frac{l(\epsilon_A-1)}{l(\epsilon_A+1)+1}\bigg{[}\frac{q_p}{(R-r)}\bigg{(}\frac{-a_A}{R-r}\bigg{)}^l  \nonumber \\
&&+\bigg{(}\frac{-a_A}{R}\bigg{)}^l\sum_{m=1}^{\infty}\bigg{(}\frac{a_B}{R}\bigg{)}^{m+1}\frac{(m+l)!}{m!l!}\delta c_m^B\bigg{]},
\end{eqnarray}
for $l>0$ ($\delta c_0^B$=$\delta c_0^A$=0). In like manner as in the preceding section these equations may be written as two matrix equations,
 \begin{eqnarray}\label{eq:delta_h_B}
\bf (h^B-I)\delta c^B&=&\bf \delta w^B \label{eq:delta_h_A}\\
\bf (h^A-I)\delta c^A&=&\bf \delta w^A,
\end{eqnarray}
where the vector elements in this case are given by
\begin{eqnarray}
 \delta w^B_{l}&=&f(\epsilon_B,l)q_p\bigg{[}\frac{1}{r}\bigg{(}\frac{a_B}{r}\bigg{)}^{l} \nonumber\\
&-&\frac{a_A}{R^2-Rr}\bigg{(}\frac{a_B}{R}\bigg{)}^{l}\sum_{j=1}^{\infty}f(\epsilon_A,j)\frac{(j+l)!}{j!l!}\bigg{(}\frac{a_A^2}{R^2-Rr}\bigg{)}^{j}\bigg{]} \nonumber \\ \\
\delta w^A_{l}&=&f(\epsilon_A,l)q_p\bigg{[}\frac{1}{R-r}\bigg{(}\frac{-a_A}{R-r}\bigg{)}^{l} \nonumber \\ &-&\frac{a_B}{Rr}\bigg{(}\frac{-a_A}{R}\bigg{)}^{l}\sum_{j=1}^{\infty}f(\epsilon_B,j)\frac{(j+l)!}{j!l!}\bigg{(}\frac{a_B^2}{Rr}\bigg{)}^{j}\bigg{]},
\end{eqnarray}
while the matrix elements ($h^B_{l,m}$ and $h^A_{l,m}$) are given by Eqs. (\ref{eq:H_B}) and (\ref{eq:H_A}), and $f(\epsilon,j)$ is given by Eq. (\ref{eq:f}).

\section{Reaction cross sections}

In classical capture theory, the collisions between two (charged) particles are described by an effective potential
\begin{equation}\label{eq:U_eff}
U_{eff}(R)=U_{int}(R)+E\frac{b^2}{R^2},
\end{equation}
which includes the particle-particle interaction energy $U_{int}(R)$, and a centrifugal term where $E$ is the initial kinetic energy available in the center of mass system, $b$ is the impact parameter, and $R$ is the distance between the particles. The second term may give rise to a centrifugal barrier, which means that the kinetic energy needs to be larger than the barrier height  ($E\geq U_{eff}(R=R_b)$) for a reaction to occur. The position (R$_b$) of the maximum barrier height may be found by solving 
\begin{equation}\label{eq:dU_eff}
\frac{dU_{eff}}{dR}=\frac{dU_{int}}{dR}-2E\frac{b^2}{R^3}=0 \rightarrow \frac{R}{2} \frac{dU_{int}}{dR}=E\frac{b^2}{R^2}.
\end{equation}
and substituting this results into Eq. \ref{eq:U_eff} 
\begin{equation}\label{eq:Rb}
E=U_{int}(R_b)+\frac{R_b}{2} \frac{dU_{int}(R_b)}{dR}.
\end{equation}
The corresponding impact parameter is then 
\begin{equation}\label{eq:b}
b=R_b\sqrt{1-U_{int}(R_b)/E}
\end{equation}
and the reaction cross section
\begin{equation}\label{eq:langevin_cross}
\sigma_{Reac}=\pi b^2= \pi R_b^2(1-U_{int}(R_b)/E).
\end{equation}
Eq. \ref{eq:dU_eff} has to be solved numerically for the interaction between two charged dielectric spheres, but it is possible to solve in the special case of a point charge interacting with a neutral sphere ($q_B$=0). The approximate expression (Eq. \ref{eq:Uint_point_approx}) for the interaction energy is then 
\begin{equation}\label{eq:Uint-neutral sphere}
U_{int}(R) \approx - \frac{\epsilon_{B}-1}{\epsilon_{B}+2}\frac{q_{A}^2a_{B}^3}{2R^2(R^2-a_{B}^2)},
\end{equation}
which gives 
\begin{equation}\label{eq:Rb_neutral}
R_b=a_B\sqrt{1+ \sqrt{\frac{\epsilon_{B}-1}{\epsilon_{B}+2}\frac{q_{A}^2}{2a_BE}}}
\end{equation}
and the reaction cross section (Eq. \ref{eq:langevin_cross})
\begin{equation}\label{eq:langevin_cross_approx}
\sigma_{Reac}=\pi a_B^2\bigg{(}1+ \sqrt{\frac{\epsilon_{B}-1}{\epsilon_{B}+2}\frac{2q_{A}^2}{a_BE}}\bigg{)}.
\end{equation}
For high energies $\sigma_{Reac}=\pi a_B^2$ (hard sphere collision) and for low energies  $\sigma_{Reac}\propto E^{-1/2}$, which means that the rate ($\sigma_{Reac}v$) is constant in the latter case.    


\begin{thebibliography}{37}%
\makeatletter
\providecommand \@ifxundefined [1]{%
 \@ifx{#1\undefined}
}%
\providecommand \@ifnum [1]{%
 \ifnum #1\expandafter \@firstoftwo
 \else \expandafter \@secondoftwo
 \fi
}%
\providecommand \@ifx [1]{%
 \ifx #1\expandafter \@firstoftwo
 \else \expandafter \@secondoftwo
 \fi
}%
\providecommand \natexlab [1]{#1}%
\providecommand \enquote  [1]{``#1''}%
\providecommand \bibnamefont  [1]{#1}%
\providecommand \bibfnamefont [1]{#1}%
\providecommand \citenamefont [1]{#1}%
\providecommand \href@noop [0]{\@secondoftwo}%
\providecommand \href [0]{\begingroup \@sanitize@url \@href}%
\providecommand \@href[1]{\@@startlink{#1}\@@href}%
\providecommand \@@href[1]{\endgroup#1\@@endlink}%
\providecommand \@sanitize@url [0]{\catcode `\\12\catcode `\$12\catcode
  `\&12\catcode `\#12\catcode `\^12\catcode `\_12\catcode `\%12\relax}%
\providecommand \@@startlink[1]{}%
\providecommand \@@endlink[0]{}%
\providecommand \url  [0]{\begingroup\@sanitize@url \@url }%
\providecommand \@url [1]{\endgroup\@href {#1}{\urlprefix }}%
\providecommand \urlprefix  [0]{URL }%
\providecommand \Eprint [0]{\href }%
\providecommand \doibase [0]{http://dx.doi.org/}%
\providecommand \selectlanguage [0]{\@gobble}%
\providecommand \bibinfo  [0]{\@secondoftwo}%
\providecommand \bibfield  [0]{\@secondoftwo}%
\providecommand \translation [1]{[#1]}%
\providecommand \BibitemOpen [0]{}%
\providecommand \bibitemStop [0]{}%
\providecommand \bibitemNoStop [0]{.\EOS\space}%
\providecommand \EOS [0]{\spacefactor3000\relax}%
\providecommand \BibitemShut  [1]{\csname bibitem#1\endcsname}%
\let\auto@bib@innerbib\@empty
\bibitem [{\citenamefont {Ochs}\ and\ \citenamefont
  {Czys}(1987)}]{Ochs:1987gsa}%
  \BibitemOpen
  \bibfield  {author} {\bibinfo {author} {\bibfnamefont {H.~T.}\ \bibnamefont
  {Ochs}}\ and\ \bibinfo {author} {\bibfnamefont {R.~R.}\ \bibnamefont
  {Czys}},\ }\href@noop {} {\bibfield  {journal} {\bibinfo  {journal} {Nature}\
  }\textbf {\bibinfo {volume} {327}},\ \bibinfo {pages} {606} (\bibinfo {year}
  {1987})}\BibitemShut {NoStop}%
\bibitem [{\citenamefont {Bowen}\ and\ \citenamefont
  {Sharif}(1998)}]{Bowen:1998ib}%
  \BibitemOpen
  \bibfield  {author} {\bibinfo {author} {\bibfnamefont {W.~R.}\ \bibnamefont
  {Bowen}}\ and\ \bibinfo {author} {\bibfnamefont {A.~O.}\ \bibnamefont
  {Sharif}},\ }\href@noop {} {\bibfield  {journal} {\bibinfo  {journal}
  {Nature}\ }\textbf {\bibinfo {volume} {393}},\ \bibinfo {pages} {663}
  (\bibinfo {year} {1998})}\BibitemShut {NoStop}%
\bibitem [{\citenamefont {Huber}\ \emph {et~al.}(2016)\citenamefont {Huber},
  \citenamefont {Gatchell}, \citenamefont {Zettergren},\ and\ \citenamefont
  {Mauracher}}]{Huber:2016}%
  \BibitemOpen
  \bibfield  {author} {\bibinfo {author} {\bibfnamefont {S.~E.}\ \bibnamefont
  {Huber}}, \bibinfo {author} {\bibfnamefont {M.}~\bibnamefont {Gatchell}},
  \bibinfo {author} {\bibfnamefont {H.}~\bibnamefont {Zettergren}}, \ and\
  \bibinfo {author} {\bibfnamefont {A.}~\bibnamefont {Mauracher}},\ }\href@noop
  {} {\bibfield  {journal} {\bibinfo  {journal} {Carbon}\ }\textbf {\bibinfo
  {volume} {109}},\ \bibinfo {pages} {843} (\bibinfo {year}
  {2016})}\BibitemShut {NoStop}%
\bibitem [{\citenamefont {Wong}\ and\ \citenamefont
  {Pollack}(2010)}]{Wong:2010kb}%
  \BibitemOpen
  \bibfield  {author} {\bibinfo {author} {\bibfnamefont {G.~C.~L.}\
  \bibnamefont {Wong}}\ and\ \bibinfo {author} {\bibfnamefont {L.}~\bibnamefont
  {Pollack}},\ }\href@noop {} {\bibfield  {journal} {\bibinfo  {journal}
  {Annual Review of Physical Chemistry, Vol 61}\ }\textbf {\bibinfo {volume}
  {61}},\ \bibinfo {pages} {171} (\bibinfo {year} {2010})}\BibitemShut
  {NoStop}%
\bibitem [{\citenamefont {Taccogna}(2012)}]{Taccogna:2012fa}%
  \BibitemOpen
  \bibfield  {author} {\bibinfo {author} {\bibfnamefont {F.}~\bibnamefont
  {Taccogna}},\ }\href@noop {} {\bibfield  {journal} {\bibinfo  {journal}
  {Contributions to Plasma Physics}\ }\textbf {\bibinfo {volume} {52}},\
  \bibinfo {pages} {744} (\bibinfo {year} {2012})}\BibitemShut {NoStop}%
\bibitem [{\citenamefont {Draine}\ and\ \citenamefont
  {Sutin}(1987)}]{Draine:1987va}%
  \BibitemOpen
  \bibfield  {author} {\bibinfo {author} {\bibfnamefont {B.~T.}\ \bibnamefont
  {Draine}}\ and\ \bibinfo {author} {\bibfnamefont {B.}~\bibnamefont {Sutin}},\
  }\href@noop {} {\bibfield  {journal} {\bibinfo  {journal} {The Astrophysical
  Journal}\ }\textbf {\bibinfo {volume} {320}},\ \bibinfo {pages} {803}
  (\bibinfo {year} {1987})}\BibitemShut {NoStop}%
\bibitem [{\citenamefont {Feng}(2000)}]{Feng:2000dc}%
  \BibitemOpen
  \bibfield  {author} {\bibinfo {author} {\bibfnamefont {J.~Q.}\ \bibnamefont
  {Feng}},\ }\href@noop {} {\bibfield  {journal} {\bibinfo  {journal} {Physical
  Review E}\ }\textbf {\bibinfo {volume} {62}},\ \bibinfo {pages} {2891}
  (\bibinfo {year} {2000})}\BibitemShut {NoStop}%
\bibitem [{\citenamefont {Eichelberger}, \citenamefont {Snow},\ and\
  \citenamefont {Bierbaum}(2003)}]{Eichelberger:2003hk}%
  \BibitemOpen
  \bibfield  {author} {\bibinfo {author} {\bibfnamefont {B.~R.}\ \bibnamefont
  {Eichelberger}}, \bibinfo {author} {\bibfnamefont {T.~P.}\ \bibnamefont
  {Snow}}, \ and\ \bibinfo {author} {\bibfnamefont {V.~M.}\ \bibnamefont
  {Bierbaum}},\ }\href@noop {} {\bibfield  {journal} {\bibinfo  {journal}
  {Journal of the American Society for Mass Spectrometry}\ }\textbf {\bibinfo
  {volume} {14}},\ \bibinfo {pages} {501} (\bibinfo {year} {2003})}\BibitemShut
  {NoStop}%
\bibitem [{\citenamefont {Kasperovich}, \citenamefont {Tikhonov},\ and\
  \citenamefont {Kresin}(2001)}]{Kasperovich200155}%
  \BibitemOpen
  \bibfield  {author} {\bibinfo {author} {\bibfnamefont {V.}~\bibnamefont
  {Kasperovich}}, \bibinfo {author} {\bibfnamefont {G.}~\bibnamefont
  {Tikhonov}}, \ and\ \bibinfo {author} {\bibfnamefont {V.}~\bibnamefont
  {Kresin}},\ }\href {\doibase http://dx.doi.org/10.1016/S0009-2614(01)00199-3}
  {\bibfield  {journal} {\bibinfo  {journal} {Chemical Physics Letters}\
  }\textbf {\bibinfo {volume} {337}},\ \bibinfo {pages} {55 } (\bibinfo {year}
  {2001})}\BibitemShut {NoStop}%
\bibitem [{\citenamefont {Kasperovich}\ \emph {et~al.}(2000)\citenamefont
  {Kasperovich}, \citenamefont {Wong}, \citenamefont {Tikhonov},\ and\
  \citenamefont {Kresin}}]{Kasperovich:2000aa}%
  \BibitemOpen
  \bibfield  {author} {\bibinfo {author} {\bibfnamefont {V.}~\bibnamefont
  {Kasperovich}}, \bibinfo {author} {\bibfnamefont {K.}~\bibnamefont {Wong}},
  \bibinfo {author} {\bibfnamefont {G.}~\bibnamefont {Tikhonov}}, \ and\
  \bibinfo {author} {\bibfnamefont {V.~V.}\ \bibnamefont {Kresin}},\ }\href
  {\doibase 10.1103/PhysRevLett.85.2729} {\bibfield  {journal} {\bibinfo
  {journal} {Phys. Rev. Lett.}\ }\textbf {\bibinfo {volume} {85}},\ \bibinfo
  {pages} {2729} (\bibinfo {year} {2000})}\BibitemShut {NoStop}%
\bibitem [{\citenamefont {Langevin}(1905)}]{Langevin:1905wk}%
  \BibitemOpen
  \bibfield  {author} {\bibinfo {author} {\bibfnamefont {M.~P.}\ \bibnamefont
  {Langevin}},\ }\href@noop {} {\bibfield  {journal} {\bibinfo  {journal}
  {Annales de Chimie et de Physique, series}\ }\textbf {\bibinfo {volume}
  {5}},\ \bibinfo {pages} {245} (\bibinfo {year} {1905})}\BibitemShut {NoStop}%
\bibitem [{\citenamefont {Su}, \citenamefont {Viggiano},\ and\ \citenamefont
  {Paulson}(1992)}]{Su:1992co}%
  \BibitemOpen
  \bibfield  {author} {\bibinfo {author} {\bibfnamefont {T.}~\bibnamefont
  {Su}}, \bibinfo {author} {\bibfnamefont {A.~A.}\ \bibnamefont {Viggiano}}, \
  and\ \bibinfo {author} {\bibfnamefont {J.~F.}\ \bibnamefont {Paulson}},\
  }\href@noop {} {\bibfield  {journal} {\bibinfo  {journal} {The Journal of
  Chemical Physics}\ }\textbf {\bibinfo {volume} {96}},\ \bibinfo {pages}
  {5550} (\bibinfo {year} {1992})}\BibitemShut {NoStop}%
\bibitem [{\citenamefont {Hsieh}\ and\ \citenamefont
  {Castleman~Jr.}(1981)}]{Hsieh:1981dn}%
  \BibitemOpen
  \bibfield  {author} {\bibinfo {author} {\bibfnamefont {E.~T.-Y.}\
  \bibnamefont {Hsieh}}\ and\ \bibinfo {author} {\bibfnamefont {A.~W.}\
  \bibnamefont {Castleman~Jr.}},\ }\href@noop {} {\bibfield  {journal}
  {\bibinfo  {journal} {International Journal of Mass Spectrometry and Ion
  Physics}\ }\textbf {\bibinfo {volume} {40}},\ \bibinfo {pages} {295}
  (\bibinfo {year} {1981})}\BibitemShut {NoStop}%
\bibitem [{\citenamefont {Su}, \citenamefont {Su},\ and\ \citenamefont
  {Bowers}(1978)}]{Su:1978gg}%
  \BibitemOpen
  \bibfield  {author} {\bibinfo {author} {\bibfnamefont {T.}~\bibnamefont
  {Su}}, \bibinfo {author} {\bibfnamefont {E.~C.~F.}\ \bibnamefont {Su}}, \
  and\ \bibinfo {author} {\bibfnamefont {M.~T.}\ \bibnamefont {Bowers}},\
  }\href@noop {} {\bibfield  {journal} {\bibinfo  {journal} {International
  Journal of Mass Spectrometry and Ion Physics}\ }\textbf {\bibinfo {volume}
  {28}},\ \bibinfo {pages} {285} (\bibinfo {year} {1978})}\BibitemShut
  {NoStop}%
\bibitem [{\citenamefont {B{\'a}r{\'a}ny}\ \emph {et~al.}(1985)\citenamefont
  {B{\'a}r{\'a}ny}, \citenamefont {Astner}, \citenamefont {Cederquist},
  \citenamefont {Danared}, \citenamefont {Huldt}, \citenamefont {Hvelplund},
  \citenamefont {Johnson}, \citenamefont {Knudsen}, \citenamefont {Liljeby},\
  and\ \citenamefont {Rensfelt}}]{Barany:1985cq}%
  \BibitemOpen
  \bibfield  {author} {\bibinfo {author} {\bibfnamefont {A.}~\bibnamefont
  {B{\'a}r{\'a}ny}}, \bibinfo {author} {\bibfnamefont {G.}~\bibnamefont
  {Astner}}, \bibinfo {author} {\bibfnamefont {H.}~\bibnamefont {Cederquist}},
  \bibinfo {author} {\bibfnamefont {H.}~\bibnamefont {Danared}}, \bibinfo
  {author} {\bibfnamefont {S.}~\bibnamefont {Huldt}}, \bibinfo {author}
  {\bibfnamefont {P.}~\bibnamefont {Hvelplund}}, \bibinfo {author}
  {\bibfnamefont {A.}~\bibnamefont {Johnson}}, \bibinfo {author} {\bibfnamefont
  {H.}~\bibnamefont {Knudsen}}, \bibinfo {author} {\bibfnamefont
  {L.}~\bibnamefont {Liljeby}}, \ and\ \bibinfo {author} {\bibfnamefont
  {K.~G.}\ \bibnamefont {Rensfelt}},\ }\href@noop {} {\bibfield  {journal}
  {\bibinfo  {journal} {Nuclear Instruments {\&} Methods in Physics Research
  Section B-Beam Interactions with Materials and Atoms}\ }\textbf {\bibinfo
  {volume} {9}},\ \bibinfo {pages} {397} (\bibinfo {year} {1985})}\BibitemShut
  {NoStop}%
\bibitem [{\citenamefont {Niehaus}(1986)}]{Niehaus:1986kg}%
  \BibitemOpen
  \bibfield  {author} {\bibinfo {author} {\bibfnamefont {A.}~\bibnamefont
  {Niehaus}},\ }\href@noop {} {\bibfield  {journal} {\bibinfo  {journal}
  {Journal of Physics B: Atomic and Molecular Physics}\ }\textbf {\bibinfo
  {volume} {19}},\ \bibinfo {pages} {2925} (\bibinfo {year}
  {1986})}\BibitemShut {NoStop}%
\bibitem [{\citenamefont {B{\'a}r{\'a}ny}\ and\ \citenamefont
  {Setterlind}(1995)}]{Barany:1995hr}%
  \BibitemOpen
  \bibfield  {author} {\bibinfo {author} {\bibfnamefont {A.}~\bibnamefont
  {B{\'a}r{\'a}ny}}\ and\ \bibinfo {author} {\bibfnamefont {C.~J.}\
  \bibnamefont {Setterlind}},\ }\href@noop {} {\bibfield  {journal} {\bibinfo
  {journal} {Nuclear Instruments {\&} Methods in Physics Research Section
  B-Beam Interactions with Materials and Atoms}\ }\textbf {\bibinfo {volume}
  {98}},\ \bibinfo {pages} {184} (\bibinfo {year} {1995})}\BibitemShut
  {NoStop}%
\bibitem [{\citenamefont {Cederquist}\ \emph {et~al.}(2000)\citenamefont
  {Cederquist}, \citenamefont {Fardi}, \citenamefont {Haghighat}, \citenamefont
  {Langereis}, \citenamefont {Schmidt}, \citenamefont {Schwartz}, \citenamefont
  {Levin}, \citenamefont {Sellin}, \citenamefont {Lebius}, \citenamefont
  {Huber}, \citenamefont {Larsson},\ and\ \citenamefont
  {Hvelplund}}]{Cederquist:2000hv}%
  \BibitemOpen
  \bibfield  {author} {\bibinfo {author} {\bibfnamefont {H.}~\bibnamefont
  {Cederquist}}, \bibinfo {author} {\bibfnamefont {A.}~\bibnamefont {Fardi}},
  \bibinfo {author} {\bibfnamefont {K.}~\bibnamefont {Haghighat}}, \bibinfo
  {author} {\bibfnamefont {A.}~\bibnamefont {Langereis}}, \bibinfo {author}
  {\bibfnamefont {H.~T.}\ \bibnamefont {Schmidt}}, \bibinfo {author}
  {\bibfnamefont {S.~H.}\ \bibnamefont {Schwartz}}, \bibinfo {author}
  {\bibfnamefont {J.~C.}\ \bibnamefont {Levin}}, \bibinfo {author}
  {\bibfnamefont {I.~A.}\ \bibnamefont {Sellin}}, \bibinfo {author}
  {\bibfnamefont {H.}~\bibnamefont {Lebius}}, \bibinfo {author} {\bibfnamefont
  {B.}~\bibnamefont {Huber}}, \bibinfo {author} {\bibfnamefont {M.~O.}\
  \bibnamefont {Larsson}}, \ and\ \bibinfo {author} {\bibfnamefont
  {P.}~\bibnamefont {Hvelplund}},\ }\href@noop {} {\bibfield  {journal}
  {\bibinfo  {journal} {Physical Review A}\ }\textbf {\bibinfo {volume} {61}},\
  \bibinfo {pages} {022712} (\bibinfo {year} {2000})}\BibitemShut {NoStop}%
\bibitem [{\citenamefont {Lawicki}\ \emph {et~al.}(2011)\citenamefont
  {Lawicki}, \citenamefont {Holm}, \citenamefont {Rousseau}, \citenamefont
  {Capron}, \citenamefont {Maisonny}, \citenamefont {Maclot}, \citenamefont
  {Seitz}, \citenamefont {Johansson}, \citenamefont {Rosen}, \citenamefont
  {Schmidt}, \citenamefont {Zettergren}, \citenamefont {Manil}, \citenamefont
  {Adoui}, \citenamefont {Cederquist},\ and\ \citenamefont
  {Huber}}]{Lawicki:2011fc}%
  \BibitemOpen
  \bibfield  {author} {\bibinfo {author} {\bibfnamefont {A.}~\bibnamefont
  {Lawicki}}, \bibinfo {author} {\bibfnamefont {A.~I.~S.}\ \bibnamefont
  {Holm}}, \bibinfo {author} {\bibfnamefont {P.}~\bibnamefont {Rousseau}},
  \bibinfo {author} {\bibfnamefont {M.}~\bibnamefont {Capron}}, \bibinfo
  {author} {\bibfnamefont {R.}~\bibnamefont {Maisonny}}, \bibinfo {author}
  {\bibfnamefont {S.}~\bibnamefont {Maclot}}, \bibinfo {author} {\bibfnamefont
  {F.}~\bibnamefont {Seitz}}, \bibinfo {author} {\bibfnamefont {H.~A.~B.}\
  \bibnamefont {Johansson}}, \bibinfo {author} {\bibfnamefont {S.}~\bibnamefont
  {Rosen}}, \bibinfo {author} {\bibfnamefont {H.~T.}\ \bibnamefont {Schmidt}},
  \bibinfo {author} {\bibfnamefont {H.}~\bibnamefont {Zettergren}}, \bibinfo
  {author} {\bibfnamefont {B.}~\bibnamefont {Manil}}, \bibinfo {author}
  {\bibfnamefont {L.}~\bibnamefont {Adoui}}, \bibinfo {author} {\bibfnamefont
  {H.}~\bibnamefont {Cederquist}}, \ and\ \bibinfo {author} {\bibfnamefont
  {B.~A.}\ \bibnamefont {Huber}},\ }\href@noop {} {\bibfield  {journal}
  {\bibinfo  {journal} {Physical Review A}\ }\textbf {\bibinfo {volume} {83}},\
  \bibinfo {pages} {022704} (\bibinfo {year} {2011})}\BibitemShut {NoStop}%
\bibitem [{\citenamefont {Forsberg}\ \emph {et~al.}(2013)\citenamefont
  {Forsberg}, \citenamefont {Alexander}, \citenamefont {Chen}, \citenamefont
  {Pettersson}, \citenamefont {Gatchell}, \citenamefont {Cederquist},\ and\
  \citenamefont {Zettergren}}]{Forsberg:2013cs}%
  \BibitemOpen
  \bibfield  {author} {\bibinfo {author} {\bibfnamefont {B.~O.}\ \bibnamefont
  {Forsberg}}, \bibinfo {author} {\bibfnamefont {J.~D.}\ \bibnamefont
  {Alexander}}, \bibinfo {author} {\bibfnamefont {T.}~\bibnamefont {Chen}},
  \bibinfo {author} {\bibfnamefont {A.~T.}\ \bibnamefont {Pettersson}},
  \bibinfo {author} {\bibfnamefont {M.}~\bibnamefont {Gatchell}}, \bibinfo
  {author} {\bibfnamefont {H.}~\bibnamefont {Cederquist}}, \ and\ \bibinfo
  {author} {\bibfnamefont {H.}~\bibnamefont {Zettergren}},\ }\href@noop {}
  {\bibfield  {journal} {\bibinfo  {journal} {The Journal of Chemical Physics}\
  }\textbf {\bibinfo {volume} {138}},\ \bibinfo {pages} {054306} (\bibinfo
  {year} {2013})}\BibitemShut {NoStop}%
\bibitem [{\citenamefont {Zettergren}\ \emph {et~al.}(2002)\citenamefont
  {Zettergren}, \citenamefont {Schmidt}, \citenamefont {Cederquist},
  \citenamefont {Jensen}, \citenamefont {Tomita}, \citenamefont {Hvelplund},
  \citenamefont {Lebius},\ and\ \citenamefont {Huber}}]{Zettergren:2002ja}%
  \BibitemOpen
  \bibfield  {author} {\bibinfo {author} {\bibfnamefont {H.}~\bibnamefont
  {Zettergren}}, \bibinfo {author} {\bibfnamefont {H.~T.}\ \bibnamefont
  {Schmidt}}, \bibinfo {author} {\bibfnamefont {H.}~\bibnamefont {Cederquist}},
  \bibinfo {author} {\bibfnamefont {J.}~\bibnamefont {Jensen}}, \bibinfo
  {author} {\bibfnamefont {S.}~\bibnamefont {Tomita}}, \bibinfo {author}
  {\bibfnamefont {P.}~\bibnamefont {Hvelplund}}, \bibinfo {author}
  {\bibfnamefont {H.}~\bibnamefont {Lebius}}, \ and\ \bibinfo {author}
  {\bibfnamefont {B.~A.}\ \bibnamefont {Huber}},\ }\href@noop {} {\bibfield
  {journal} {\bibinfo  {journal} {Physical Review A}\ }\textbf {\bibinfo
  {volume} {66}},\ \bibinfo {pages} {032710} (\bibinfo {year}
  {2002})}\BibitemShut {NoStop}%
\bibitem [{\citenamefont {Burgd{\"o}rfer}, \citenamefont {Lerner},\ and\
  \citenamefont {Meyer}(1991)}]{Burgdorfer:1991gc}%
  \BibitemOpen
  \bibfield  {author} {\bibinfo {author} {\bibfnamefont {J.}~\bibnamefont
  {Burgd{\"o}rfer}}, \bibinfo {author} {\bibfnamefont {P.}~\bibnamefont
  {Lerner}}, \ and\ \bibinfo {author} {\bibfnamefont {F.~W.}\ \bibnamefont
  {Meyer}},\ }\href@noop {} {\bibfield  {journal} {\bibinfo  {journal}
  {Physical Review A}\ }\textbf {\bibinfo {volume} {44}},\ \bibinfo {pages}
  {5674} (\bibinfo {year} {1991})}\BibitemShut {NoStop}%
\bibitem [{\citenamefont {Wethekam}\ \emph {et~al.}(2007)\citenamefont
  {Wethekam}, \citenamefont {Winter}, \citenamefont {Cederquist},\ and\
  \citenamefont {Zettergren}}]{Wethekam:2007be}%
  \BibitemOpen
  \bibfield  {author} {\bibinfo {author} {\bibfnamefont {S.}~\bibnamefont
  {Wethekam}}, \bibinfo {author} {\bibfnamefont {H.}~\bibnamefont {Winter}},
  \bibinfo {author} {\bibfnamefont {H.}~\bibnamefont {Cederquist}}, \ and\
  \bibinfo {author} {\bibfnamefont {H.}~\bibnamefont {Zettergren}},\
  }\href@noop {} {\bibfield  {journal} {\bibinfo  {journal} {Physical Review
  Letters}\ }\textbf {\bibinfo {volume} {99}},\ \bibinfo {pages} {037601}
  (\bibinfo {year} {2007})}\BibitemShut {NoStop}%
\bibitem [{\citenamefont {Zettergren}, \citenamefont {Forsberg},\ and\
  \citenamefont {Cederquist}(2012)}]{Zettergren:2012ft}%
  \BibitemOpen
  \bibfield  {author} {\bibinfo {author} {\bibfnamefont {H.}~\bibnamefont
  {Zettergren}}, \bibinfo {author} {\bibfnamefont {B.~O.}\ \bibnamefont
  {Forsberg}}, \ and\ \bibinfo {author} {\bibfnamefont {H.}~\bibnamefont
  {Cederquist}},\ }\href@noop {} {\bibfield  {journal} {\bibinfo  {journal}
  {Physical Chemistry Chemical Physics}\ }\textbf {\bibinfo {volume} {14}},\
  \bibinfo {pages} {16360} (\bibinfo {year} {2012})}\BibitemShut {NoStop}%
\bibitem [{\citenamefont {Lindell}, \citenamefont {Sten},\ and\ \citenamefont
  {Nikoskinen}(1993)}]{Lindell:1993jy}%
  \BibitemOpen
  \bibfield  {author} {\bibinfo {author} {\bibfnamefont {I.~V.}\ \bibnamefont
  {Lindell}}, \bibinfo {author} {\bibfnamefont {J.~C.~E.}\ \bibnamefont
  {Sten}}, \ and\ \bibinfo {author} {\bibfnamefont {K.~I.}\ \bibnamefont
  {Nikoskinen}},\ }\href@noop {} {\bibfield  {journal} {\bibinfo  {journal}
  {Radio science}\ }\textbf {\bibinfo {volume} {28}},\ \bibinfo {pages} {319}
  (\bibinfo {year} {1993})}\BibitemShut {NoStop}%
\bibitem [{\citenamefont {Bichoutskaia}\ \emph {et~al.}(2010)\citenamefont
  {Bichoutskaia}, \citenamefont {Boatwright}, \citenamefont {Khachatourian},\
  and\ \citenamefont {Stace}}]{Bichoutskaia:2010bf}%
  \BibitemOpen
  \bibfield  {author} {\bibinfo {author} {\bibfnamefont {E.}~\bibnamefont
  {Bichoutskaia}}, \bibinfo {author} {\bibfnamefont {A.~L.}\ \bibnamefont
  {Boatwright}}, \bibinfo {author} {\bibfnamefont {A.}~\bibnamefont
  {Khachatourian}}, \ and\ \bibinfo {author} {\bibfnamefont {A.~J.}\
  \bibnamefont {Stace}},\ }\href@noop {} {\bibfield  {journal} {\bibinfo
  {journal} {The Journal of Chemical Physics}\ }\textbf {\bibinfo {volume}
  {133}},\ \bibinfo {pages} {024105} (\bibinfo {year} {2010})}\BibitemShut
  {NoStop}%
\bibitem [{\citenamefont {Nakajima}\ and\ \citenamefont
  {Sato}(1999)}]{Nakajima:1999cr}%
  \BibitemOpen
  \bibfield  {author} {\bibinfo {author} {\bibfnamefont {Y.}~\bibnamefont
  {Nakajima}}\ and\ \bibinfo {author} {\bibfnamefont {T.}~\bibnamefont
  {Sato}},\ }\href@noop {} {\bibfield  {journal} {\bibinfo  {journal} {Journal
  of electrostatics}\ }\textbf {\bibinfo {volume} {45}},\ \bibinfo {pages}
  {213} (\bibinfo {year} {1999})}\BibitemShut {NoStop}%
\bibitem [{\citenamefont {Burgd{\"o}rfer}\ \emph {et~al.}(1996)\citenamefont
  {Burgd{\"o}rfer}, \citenamefont {Reinhold}, \citenamefont {H{\"a}gg},\ and\
  \citenamefont {Meyer}}]{burgdorfer96}%
  \BibitemOpen
  \bibfield  {author} {\bibinfo {author} {\bibfnamefont {J.}~\bibnamefont
  {Burgd{\"o}rfer}}, \bibinfo {author} {\bibfnamefont {C.}~\bibnamefont
  {Reinhold}}, \bibinfo {author} {\bibfnamefont {L.}~\bibnamefont {H{\"a}gg}},
  \ and\ \bibinfo {author} {\bibfnamefont {F.}~\bibnamefont {Meyer}},\
  }\href@noop {} {\bibfield  {journal} {\bibinfo  {journal} {Australian Journal
  of Physics}\ }\textbf {\bibinfo {volume} {49}},\ \bibinfo {pages} {527}
  (\bibinfo {year} {1996})}\BibitemShut {NoStop}%
\bibitem [{\citenamefont {H{\"a}gg}, \citenamefont {Reinhold},\ and\
  \citenamefont {Burgd{\"o}rfer}(1997)}]{Hagg:1997aa}%
  \BibitemOpen
  \bibfield  {author} {\bibinfo {author} {\bibfnamefont {L.}~\bibnamefont
  {H{\"a}gg}}, \bibinfo {author} {\bibfnamefont {C.~O.}\ \bibnamefont
  {Reinhold}}, \ and\ \bibinfo {author} {\bibfnamefont {J.}~\bibnamefont
  {Burgd{\"o}rfer}},\ }\href {\doibase 10.1103/PhysRevA.55.2097} {\bibfield
  {journal} {\bibinfo  {journal} {Physical Review A}\ }\textbf {\bibinfo
  {volume} {55}},\ \bibinfo {pages} {2097} (\bibinfo {year}
  {1997})}\BibitemShut {NoStop}%
\bibitem [{\citenamefont {Jackson}(1975)}]{jackson75}%
  \BibitemOpen
  \bibfield  {author} {\bibinfo {author} {\bibfnamefont {J.~D.}\ \bibnamefont
  {Jackson}},\ }\href@noop {} {\emph {\bibinfo {title} {{C}lassical
  {E}lectrodynamics}}},\ \bibinfo {edition} {2nd}\ ed.\ (\bibinfo  {publisher}
  {John Wiley {\&} Sons},\ \bibinfo {year} {1975})\BibitemShut {NoStop}%
\bibitem [{\citenamefont {Zettergren}\ \emph {et~al.}(2007)\citenamefont
  {Zettergren}, \citenamefont {Schmidt}, \citenamefont {Reinhed}, \citenamefont
  {Cederquist}, \citenamefont {Jensen}, \citenamefont {Hvelplund},
  \citenamefont {Tomita}, \citenamefont {Manil}, \citenamefont {Rangama},\ and\
  \citenamefont {Huber}}]{Zettergren:2007ig}%
  \BibitemOpen
  \bibfield  {author} {\bibinfo {author} {\bibfnamefont {H.}~\bibnamefont
  {Zettergren}}, \bibinfo {author} {\bibfnamefont {H.~T.}\ \bibnamefont
  {Schmidt}}, \bibinfo {author} {\bibfnamefont {P.}~\bibnamefont {Reinhed}},
  \bibinfo {author} {\bibfnamefont {H.}~\bibnamefont {Cederquist}}, \bibinfo
  {author} {\bibfnamefont {J.}~\bibnamefont {Jensen}}, \bibinfo {author}
  {\bibfnamefont {P.}~\bibnamefont {Hvelplund}}, \bibinfo {author}
  {\bibfnamefont {S.}~\bibnamefont {Tomita}}, \bibinfo {author} {\bibfnamefont
  {B.}~\bibnamefont {Manil}}, \bibinfo {author} {\bibfnamefont
  {J.}~\bibnamefont {Rangama}}, \ and\ \bibinfo {author} {\bibfnamefont
  {B.~A.}\ \bibnamefont {Huber}},\ }\href@noop {} {\bibfield  {journal}
  {\bibinfo  {journal} {The Journal of Chemical Physics}\ }\textbf {\bibinfo
  {volume} {126}},\ \bibinfo {pages} {224303} (\bibinfo {year}
  {2007})}\BibitemShut {NoStop}%
\bibitem [{\citenamefont {{Sunil K Sainis}}\ \emph {et~al.}(2007)\citenamefont
  {{Sunil K Sainis}}, \citenamefont {{Vincent Germain}}, \citenamefont
  {Mejean}, ,\ and\ \citenamefont {Dufresne}}]{SunilKSainis:2007cp}%
  \BibitemOpen
  \bibfield  {author} {\bibinfo {author} {\bibnamefont {{Sunil K Sainis}}},
  \bibinfo {author} {\bibnamefont {{Vincent Germain}}}, \bibinfo {author}
  {\bibfnamefont {C.~O.}\ \bibnamefont {Mejean}}, , \ and\ \bibinfo {author}
  {\bibfnamefont {E.~R.}\ \bibnamefont {Dufresne}},\ }\href@noop {} {\bibfield
  {journal} {\bibinfo  {journal} {Langmuir}\ }\textbf {\bibinfo {volume}
  {24}},\ \bibinfo {pages} {1160} (\bibinfo {year} {2007})}\BibitemShut
  {NoStop}%
\bibitem [{\citenamefont {Sainis}, \citenamefont {Merrill},\ and\ \citenamefont
  {Dufresne}(2008)}]{Sainis:2008gj}%
  \BibitemOpen
  \bibfield  {author} {\bibinfo {author} {\bibfnamefont {S.~K.}\ \bibnamefont
  {Sainis}}, \bibinfo {author} {\bibfnamefont {J.~W.}\ \bibnamefont {Merrill}},
  \ and\ \bibinfo {author} {\bibfnamefont {E.~R.}\ \bibnamefont {Dufresne}},\
  }\href@noop {} {\bibfield  {journal} {\bibinfo  {journal} {Langmuir}\
  }\textbf {\bibinfo {volume} {24}},\ \bibinfo {pages} {13334} (\bibinfo {year}
  {2008})}\BibitemShut {NoStop}%
\bibitem [{\citenamefont {Hvelplund}\ \emph {et~al.}(1985)\citenamefont
  {Hvelplund}, \citenamefont {Andersen}, \citenamefont {B{\'a}r{\'a}ny},
  \citenamefont {Cederquist}, \citenamefont {Heinemeier}, \citenamefont
  {Knudsen}, \citenamefont {MacAdam}, \citenamefont {Nielsen},\ and\
  \citenamefont {S{\o}rensen}}]{hvelplund1985energy}%
  \BibitemOpen
  \bibfield  {author} {\bibinfo {author} {\bibfnamefont {P.}~\bibnamefont
  {Hvelplund}}, \bibinfo {author} {\bibfnamefont {L.}~\bibnamefont {Andersen}},
  \bibinfo {author} {\bibfnamefont {A.}~\bibnamefont {B{\'a}r{\'a}ny}},
  \bibinfo {author} {\bibfnamefont {H.}~\bibnamefont {Cederquist}}, \bibinfo
  {author} {\bibfnamefont {J.}~\bibnamefont {Heinemeier}}, \bibinfo {author}
  {\bibfnamefont {H.}~\bibnamefont {Knudsen}}, \bibinfo {author} {\bibfnamefont
  {K.}~\bibnamefont {MacAdam}}, \bibinfo {author} {\bibfnamefont
  {E.}~\bibnamefont {Nielsen}}, \ and\ \bibinfo {author} {\bibfnamefont
  {J.}~\bibnamefont {S{\o}rensen}},\ }\href@noop {} {\bibfield  {journal}
  {\bibinfo  {journal} {Nuclear Instruments and Methods in Physics Research
  Section B: Beam Interactions with Materials and Atoms}\ }\textbf {\bibinfo
  {volume} {9}},\ \bibinfo {pages} {421} (\bibinfo {year} {1985})}\BibitemShut
  {NoStop}%
\bibitem [{\citenamefont {Weingartner}\ and\ \citenamefont
  {Draine}(2001)}]{weingartner2001electron}%
  \BibitemOpen
  \bibfield  {author} {\bibinfo {author} {\bibfnamefont {J.~C.}\ \bibnamefont
  {Weingartner}}\ and\ \bibinfo {author} {\bibfnamefont {B.}~\bibnamefont
  {Draine}},\ }\href@noop {} {\bibfield  {journal} {\bibinfo  {journal} {The
  Astrophysical Journal}\ }\textbf {\bibinfo {volume} {563}},\ \bibinfo {pages}
  {842} (\bibinfo {year} {2001})}\BibitemShut {NoStop}%
\bibitem [{\citenamefont {Omont}(2016)}]{oomont2016interstellar}%
  \BibitemOpen
  \bibfield  {author} {\bibinfo {author} {\bibfnamefont {A.}~\bibnamefont
  {Omont}},\ }\href@noop {} {\bibfield  {journal} {\bibinfo  {journal}
  {Astronomy \& Astrophysics}\ }\textbf {\bibinfo {volume} {590}},\ \bibinfo
  {pages} {A52} (\bibinfo {year} {2016})}\BibitemShut {NoStop}%
\bibitem [{\citenamefont {Franklin}(2005)}]{franklin05}%
  \BibitemOpen
  \bibfield  {author} {\bibinfo {author} {\bibfnamefont {J.}~\bibnamefont
  {Franklin}},\ }\href@noop {} {\emph {\bibinfo {title} {Classical
  Electromagnetism}}}\ (\bibinfo  {publisher} {Addison-Wesley},\ \bibinfo
  {year} {2005})\BibitemShut {NoStop}%
\end{thebibliography}%
\end{document}